\documentclass[review,11pt]{elsarticle}

\usepackage{amssymb,amsmath,amsfonts,amsxtra,amsthm}
\usepackage{graphicx}
\usepackage[english]{babel}
\usepackage[utf8]{inputenc}
\usepackage[T1]{fontenc}
\usepackage{setspace}
\usepackage{undertilde}
\usepackage{lscape}
\usepackage{multirow}
\usepackage{datetime}
\usepackage{mathrsfs}
\usepackage{enumitem}
\usepackage{tabularx}
\usepackage{slashbox}
\usepackage{caption}
\usepackage[table]{xcolor}
\usepackage{algpseudocode}
\usepackage{float}
\usepackage{xcolor}

\usdate

\RequirePackage[OT1]{fontenc}
\RequirePackage[colorlinks]{hyperref}
\hypersetup{
    colorlinks=true,%
    citecolor=blue,%
    filecolor=blue,%
    linkcolor=blue,%
    urlcolor=blue
}

\theoremstyle{plain}

\newtheorem{theorem}{Theorem}

\newtheorem{corollary}[theorem]{Corollary}

\newtheorem{example}[theorem]{Example}

\newtheorem{lemma}[theorem]{Lemma}

\newtheorem{remark}[theorem]{Remark}

\newtheorem{asu}{Assumption}

\newtheorem{Algorithm}{Algorithm}

\DeclareMathOperator*{\plim}{plim}

\DeclareMathOperator*{\Var}{Var}
\DeclareMathOperator*{\argmin}{argmin}
\DeclareMathOperator*{\argsup}{argsup}
\DeclareMathOperator*{\diag}{diag}
\DeclareMathOperator*{\vec2}{vec}

\newcounter{Aequ}
\newcounter{Aaux}

\RequirePackage[OT1]{fontenc}

\definecolor{mygray}{gray}{0.75}

\usepackage{numcompress}\biboptions{authoryear}

\begin{document}

\begin{frontmatter}

\title{\textbf{A Higher-Order Correct Fast Moving-Average Bootstrap \\ for Dependent Data}}

\author[mymainaddress]{Davide La Vecchia} 

\author[mymainaddress]{Alban Moor}

\author[mysecondaryaddress]{Olivier Scaillet\corref{mycorrespondingauthor}}
\cortext[mycorrespondingauthor]{Corresponding author}
\ead{olivier.scaillet@unige.ch}

\address[mymainaddress]{Research Center for Statistics, Geneva School of Economics and Management, University of Geneva,\\ Bd Pont-d’Arve 40, CH-1211 Geneva 4, Switzerland.}
\address[mysecondaryaddress]{Geneva Finance Research Institute, University of Geneva, and Swiss Finance Institute,\\ Bd Pont-d’Arve 40, CH-1211 Geneva 4, Switzerland.}

\begin{abstract}
We develop theory of a novel fast bootstrap for dependent data. Our scheme deploys
i.i.d.\ resampling of smoothed moment indicators.
We characterize the class of parametric and semiparametric estimation problems for which the method is valid. 
We show the asymptotic refinements of the new procedure, proving that it is higher-order correct under mild assumptions on the 
time series, the estimating functions, and the smoothing kernel. We illustrate the applicability and the advantages of our 
procedure for M-estimation, generalized method of moments, and generalized empirical likelihood estimation. In a Monte Carlo study, we consider an 
autoregressive conditional duration model and we compare our method with other extant, routinely-applied first- and higher-order correct methods. The results
provide numerical evidence that the novel bootstrap yields higher-order accurate confidence intervals, while remaining computationally lighter than its higher-order correct competitors. A real-data example on 
dynamics of trading volume of US stocks illustrates the empirical relevance of our method.
\bigskip


\medskip

\end{abstract}


\end{frontmatter}

\newpage

\section{Introduction} 

Inference based on first-order correct asymptotics can be misleading with confidence intervals having erratic probability coverage. It is especially true in the presence of serial dependence where first-order asymptotics often requires larger sample sizes than for i.i.d.\ data to apply. Resampling methods for time series help to obtain confidence intervals with better finite sample properties. Bootstrap methods for moment condition models have been extensively discussed under various dependence structures by, for example, \cite{hall_bootstrap_1996}, \cite{brown_generalized_2002}, \cite{inoue_bootstrapping_2006}, and \cite{davidson_bootstrap_2006}. If bootstrap methods for $m$-dependent and strongly mixing data can achieve higher-order correctness (\cite{hall_bootstrap_1996}, \cite{inoue_bootstrapping_2006}), they are computationally too intensive, once applied to heavy numerical estimation procedures. For a book-length review, see e.g.\ \cite{lahiri_resampling_2010}.

In this paper, we propose a novel fast bootstrap scheme, that we call the Fast Moving-average Bootstrap (FMB). The resampling method is computationally attractive while maintaining higher-order correctness of the inferential procedure for strongly mixing data. Our idea for building confidence regions for the parameter of interest is to realize that smoothing the moment indicators as in the Generalized Empirical Likelihood (GEL) literature permits to bootstrap them as if they were i.i.d.\ \cite{parente_generalised_2018a} study the first-order validity of GEL test statistics based on a similar bootstrapping scheme, the Kernel Block Bootstrap (henceforth KBB); see \cite{parente_kernel_2018b} and \cite{parente_quasi-maximum_2019}. Our approach differs from KBB in two significant aspects. First, our methodology does not require to solve the estimation problem at each bootstrap sample, lessening drastically the computational burden. Indeed, FMB is at least (except for a simple low-dimensional linear model) one thousand times faster, according to standard rules on bootstrap simulation errors (\cite{efron_better_1987} Section 9, \cite{davison_bootstrap_1997} Section 2.5.2). Second, we exploit an inversion technique to benefit from the kernel smoothing used in the studentization of our test statistic. The inversion is related to the standard percentile$-t$ bootstrap approach in the linear univariate case (see Example 1 below). The studentization relies on a simple sample variance of the smoothed moment
indicators, which turns out to be asymptotically equivalent to a HAC estimator for the
original moment indicators, as shown by \cite{smith_automatic_2005}.
Together these inversion and studentization make our FMB inference amenable to be shown higher-order correct. Our proof strategy is not directly applicable to KBB; its higher-order correctness remains a conjecture.

The already existing fast resampling methods usually hinge on the first-order von Mises expansion of the estimating function (\cite{shao_jackknife_1995}, \cite{davidson_bootstrap_1999}, \cite{andrews_higher-order_2002}, \cite{salibian-barrera_bootstrapping_2002}, \cite{goncalves_maximum_2004}, \cite{hong_fast_2006}, \cite{salibian-barrera_principal_2006}, \cite{salibian-barrera_fast_2008}, \cite{camponovo_robust_2012}, \cite{camponovo_predictability_2013}, \cite{armstrong_fast_2014}, and \cite{goncalves_bootstrapping_2019}).
 It yields a fast approximation, but its inherent construction does not ensure higher-order correctness. Instead, our fast method relies on inversion, namely we identify the level sets of test statistics under the null hypothesis to obtain confidence regions for the parameter of interest (see \cite{parzen_resampling_1994} and \cite{hu_estimating_2000} for i.i.d.\ data). Furthermore, the FMB confidence regions are invariant to monotonic reparameterization, due to studentization of the moment indicators. It ensures stability of our method across varying parameter scales (\cite{diciccio_bootstrap_1996}).

We design FMB for GEL estimator to exploit its intrinsic smoothing, and as it provides a considerably wide theoretical framework on semiparametric estimation (\cite{smith_gel_2011}). As a consequence, the higher-order refinements achieved by our method ensue for the Empirical Likelihood (see \cite{qin_empirical_1994}, \cite{imbens_one-step_1996}, \cite{kitamura_empirical_1997}), the Exponential Tilting (\cite{kitamura_information-theoretic_1997}, \cite{imbens_information_1998}), and the Continuously Updating Estimator (\cite{hansen_finite-sample_1996}). In addition to the KBB, other bootstrap methods already exist in the GEL literature. For instance, \cite{bravo_empirical_2004} shows the higher-order correctness of the bootstrap for inference based on empirical likelihood with i.i.d.\ data, while \cite{bravo_blockwise_2005} shows consistency of the block bootstrap for empirical entropy tests in times series regressions with strongly mixing data. However, to our knowledge, there is no proof of higher-order correctness of the bootstrap for GEL in the literature yet.

Clearly, we can also apply FMB in the setting of M-estimation (\cite{huber_robust_1964}) and Generalized Method of Moment (\cite{hansen_large_1982}), obtaining a fast version of the bootstrap methods derived in \cite{hall_bootstrap_1996} for $m$-dependent data and \cite{inoue_bootstrapping_2006} for strongly mixing data.

The structure of the paper is as follows. Section \ref{Sketch_Methodology} is a simple introduction to the FMB algorithm in the univariate case.
  There, we also discuss connections between FMB and already existing resampling schemes. In Section \ref{MParam}, we briefly present the GMM and GEL estimators for strongly mixing time series, using these frameworks as a tool to extend FMB to the multivariate setting. We itemize our assumptions and present the main theoretical results in Section \ref{Theory}.  In Section \ref{implement}, we give details on the implementation aspects of FMB, emphasizing the relation between the choice of the kernel and the properties of the long-run variance estimator. We also discuss connections with the recent literature on confidence distributions, that we use in our empirical application. We present our Monte Carlo experiments in Section \ref{Monte_Carlo}, and a real data example in Section \ref{empirical_application}. Finally, we prove our theorems in appendix. For some technical lemmas, we give the proofs in the Supplementary Material (available online).

\section{FMB methodology}\label{method}

\subsection{An introduction in the univariate case}\label{Sketch_Methodology}

Let $\left\{X_t\right\}_{t \in \mathbb{Z}}$ be a stationary strongly mixing process in $\mathbb{R}^d$, observed at $t = 1,...,T$. We assume that the time series of interest satisfies Assumptions \ref{a1}---\ref{EE5} in Section \ref{Theory}, which are standard in the bootstrap literature. Let $\mathcal{B} 
\subset \mathbb{R}$ be the compact space of the parameter $\beta$ and $\mathbb{X}_t := \left\{ X_{t_1},...,X_{t_v} \right\}$ be a collection of vectors from the process $\left\{X_t\right\}_{t \in \mathbb{Z}}$. Consider the function $g: \mathbb{R}^{d v} \times \mathcal{B} \rightarrow \mathbb{R}$ such that:
\begin{align}
\mathbb { E } \left[ g \left( \mathbb{X} _ { t } , \beta _ { 0 } \right) \right] = 0, \label{moment}
\end{align}
where the expectation $\mathbb { E }$ is taken w.r.t.\ the true underlying distribution, unknown and depending on $\beta_0$. In the following, we use the shorthand notation $g_{t} \left( \beta \right) := g\left( \mathbb{X} _ {t}, \beta \right)$. 

\noindent The function $g$ in \eqref{moment} can be the (conditional) likelihood in full parametric models, or it can be obtained using the (conditional) moments and/or may depend on instrumental variables in semiparametric models. The collection of vectors $\mathbb{X}_t$ typically contains information on the relation between the observations and the parameter characterizing the $q$-dimensional stationary distribution of a time series. More generally, we can exploit the knowledge in closed-form of the (conditional) moments to obtain (martingale) estimating functions for non-linear conditional autoregressive and  heteroscedastic 
models or discretely observed diffusions. We refer 
to \cite{godambe_quasi-likelihood_1987}, \cite{taniguchi_asymptotic_2000}, and \cite{kessler2012statistical} for  
book-length presentations.

Each function of the sequence $\left\{ g_{t}\left( \beta \right)\right\}_{t=1}^{T}$ is often defined using the innovations, which can be i.i.d.\ random variables or more generally martingale differences. Thus, $\left\{ g_{t}\left( \beta \right)\right\}_{t=1}^{T}$ exhibits less dependence than the original process $\left\{ X_{t} \right\}_{t=1}^{T}$. Nevertheless, neglecting this temporal dependence has a serious impact on the performance of several inferential procedures, in particular it can affect the consistency of the bootstrap variance estimator and the higher-order accuracy of bootstrap confidence intervals.

To take automatically this aspect into account, we follow \cite{kitamura_information-theoretic_1997}, \cite{otsu_generalized_2006}, \cite{guggenberger_generalized_2008}, and \cite{smith_gel_2011}, and we perform 
a convolution of the moment indicator $g$ with the kernel $k: \mathbb{R} \rightarrow \mathbb{R}$, obtaining:
\begin{align}
g_{T,t}\left(\beta\right) := {B_{T}}^{-1/2}\displaystyle\sum_{s = t-T}^{t-1} k\left(\frac{s}{B_{T}}\right) g_{t-s}\left(\beta\right), \label{convol}
\end{align}
\noindent  where $B_T$ is a bandwidth parameter, increasing in $T$ and such that $B_T / T \longrightarrow 0$. The convolution in \eqref{convol} induces a HAC-type modification, ensuring consistency of the long-run variance estimation of the mean over time 
$\bar { g }_T ( \beta ) := T ^ { - 1 } \sum _ { t = 1 } ^ { T } g _ {T,t } ( \beta )$; see \cite{newey_simple_1987}, \cite{andrews_heteroscedasticity_1991}, and \cite{smith_automatic_2005}. Solving $\bar { g }_T ( \beta )=0$ gives the just-identified univariate estimator $\hat{\beta}$. Hence, the estimator $\hat{\beta}$ relies on a smoothed  moment condition. Below, we explain how we can further exploit the convolution in \eqref{convol} 
to derive our bootstrap. 

Let us first give the intuition of our methodology for the construction of confidence interval (CI) for $\beta_0$; more
technical aspects are available in Sections \ref{MParam} and \ref{sectionHAC}. 
For ease of notation, we drop the subscript $T$ from any estimator, whenever its dependence on the sample size is clear from the context.

To keep the exposition as simple as possible, we assume temporarily a one-to-one relationship between the 
parameter and the estimating function $\bar { g }_T ( \beta ),$ in an suitable subset of $\mathcal{B}$. 
Even though the probabilistic validity of the FMB CI does not depend on this condition (see e.g. \cite{lehmann_testing_1959}, \cite{shao_mathematical_1999}, \cite{hansen_finite-sample_1996} and \cite{guggenberger_generalized_2008-1}), this assumption allows us to explain easily why our resampling scheme does not need to solve the estimating equation for each bootstrap sample.

Intuitively, the construction of the FMB CI goes as follows. First, our bootstrap scheme provides a higher-order correct approximation of the distribution of a statistic $\hat{S}(\beta_0)$. This statistic is an asymptotically pivotal version of the estimating function $T^{1/2}\bar { g }_T ( \beta ),$ evaluated at the true parameter $\beta_0.$ Second, the one-to-one relationship allows us to map the quantile estimates of $\hat{S}$ to CI limits in $\mathcal{B}.$ This mapping is crucial to gain computational efficiency. Indeed, we use the computationally intensive part of the FMB algorithm to compute the distribution of simple mean-type statistic $\hat{S}(\beta_0)$ which is much faster to compute than roots of $T^{1/2}\bar { g }_T ( \beta ),$ or numerical solutions to the estimating optimization problem (see Section \ref{MParam}). From an hypothesis testing point of view, FMB yields an approximation to the distribution of $\hat S(\beta)$ under $H_0:$ $\beta=\beta_0.$ Then, each $\beta$ in the CI is in the non-rejection region of $H_0$.

The next example is a widely-applied model where the one-to-one condition is satisfied, since the considered function is monotonic. Below, we explain in Remark 2 how to adapt FMB to deal with general estimating functions, which do not necessarily satisfy the monotonicity condition.
\begin{example}\label{exAR1}
We consider an $AR(1)$ process $\{Y_t\}_{t=1}^{T},$ $Y_t = \beta Y_{t-1} + \varepsilon_t,$  where $\mid \beta \mid < 1,$ and $\{\varepsilon_t\}_{t=1}^{T}$ is a white noise. The orthogonality of the innovations yields the moment indicators $g_t(\beta)=(Y_t-\beta Y_{t-1})Y_{t-1}.$ For a given kernel $k$, smoothing these moment indicators leads to
\begin{align}
g_{T,t}(\beta) = {B_{T}}^{-1/2}\displaystyle\sum_{s = t-T}^{t-1} k\left(\frac{s}{B_{T}}\right) Y_{t-s}Y_{t-s-1} -\beta {B_{T}}^{-1/2}\displaystyle\sum_{s = t-T}^{t-1} k\left(\frac{s}{B_{T}}\right)Y_{t-s-1}^2. \label{gTtAR}
\end{align}
Equation \eqref{gTtAR} is linear in the parameter of interest $\beta.$ Thus, neither taking the mean $T^{1/2}\bar{g}_T(\beta)$ nor rescaling $T^{1/2}\bar{g}_T(\beta)$ by a constant affect this linearity. As we build our statistic of interest by rescaling $T^{1/2}\bar{g}_T(\beta),$ the one-to-one condition is verified.
\end{example}

Now that the main principles of FMB are settled, we present the detailed algorithm underlying its numerical implementation. The statistic serving as basis for inference is the asymptotically pivotal quantity:
\begin{equation}
\hat{S} := T^{1/2}\bar{g}_T\left(\beta_0\right) / \hat{\sigma}, \label{Eq S_psi}
\end{equation}
where, for instance, 
$
\hat{\sigma} := \kappa^2_1 (T\kappa_2)^{-1} \sum_{t=1}^{T} g^{2}_{T,t}( \hat{\beta} )
$ and $\kappa_j := \int k\left(u\right)^j du$ for $j=1,2$. This statistic is a particular value of the function $\hat{S}: \mathbb{R}^{d v} \times \mathcal{B} \rightarrow \mathbb{R},$ $\hat{S}(\beta):=T^{1/2}\bar{g}_T\left(\beta\right) / \hat{\sigma},$ that we suppose strictly increasing on $\mathcal{B}$. The studentization in $\hat{S}$ is crucial for FMB to be higher-order correct. In principle, we can apply other estimators of the long-run variance and we flag that each estimator $\hat{\sigma}$ has its own bias, which is going to affect the properties (e.g.\ the accuracy) of FMB. We refer to Section \ref{sectionHAC} for further discussion.

Considering an i.i.d.\ bootstrap sample drawn from $\{g_{T,t}( \hat{\beta})\}_{t=1}^{T},$ say $\{g^{\ast}_{T,t}\}_{t=1}^{T},$ the bootstrap version of $\hat{S}$ in \eqref{Eq S_psi} is: 
\begin{equation}
S^{\ast}:= T^{1/2}\bar{g}_T^\ast / \hat{\sigma}^{\ast},
\label{Eq S_psi_ast}
\end{equation}
with $\bar{g}_T^\ast:=T^{-1} \sum_{t=1}^{T} g^{\ast}_{T,t}$ and $\hat{\sigma}^{\ast 2} := {T}^{-1}\sum_{t=1}^{T} g^{\ast 2}_{T,t}.$ In \eqref{Eq S_psi_ast}, both the computed numerator and denominator rely on $g^{\ast}_{T,t},$ and thus avoid re-estimating the parameter on each bootstrap sample in order to make it fast.

Then, the algorithm of our FMB is made of five steps (lines 2-4, 5, 6-13, 14-15, 16-17). 
\begin{Algorithm}\label{AlgFMB}
\textnormal{\vspace{0.01cm}}
    \begin{algorithmic}[1] 
            \Procedure{FMB}{$X_1, ..., X_T,$  $g,$ $\alpha$} \Comment{The FMB $(1-\alpha)$-CI for $\beta_0 \in \mathcal{B}$ s.t. $\mathbb { E } \left[ g \left( \mathbb{X} _ { t } , \beta _ { 0 } \right) \right] = 0.$}
            \For{$t = 1, ..., T$}
						    \State $g_{T,t}\left(\beta\right) \leftarrow {B_{T}}^{-1/2}\sum_{s = t-T}^{t-1} k\left(s/B_{T}\right) g_{t-s}(\beta)$
						\EndFor\label{AlgFMBconvol}
						\State \textbf{Solve}(\hspace{0.05cm}$\sum_{t=1}^T g_{T,t}( \beta) = 0$) $\rightarrow$ $\hat{\beta}$
						\For{$r = 1, ..., R$}
						
						    \For{$t = 1, ..., T$}
						         \State $g_{T,t}^{\ast} \leftarrow$ \textbf{Draw}($g_{T,1}(\hat{\beta}), ..., g_{T,T}(\hat{\beta})$)
						    \EndFor\label{AlgFMBSample}
						    \State $\bar{g}_{T,r}^\ast \leftarrow T^{-1} \sum_{t=1}^{T} g^{\ast}_{T,t}$
						    \State $\hat{\sigma}_r^{\ast} \leftarrow ({T}^{-1}\sum_{t=1}^{T} g^{\ast 2}_{T,t})^{1/2}$
						    \State $S^{\ast}_r \leftarrow T^{1/2}\bar{g}_{T,r}^\ast \hat{\sigma}_r^{\ast-1}$
						\EndFor\label{AlgFMBBoot}
						\State ConfidenceLevel $\leftarrow$ $1-\alpha$
						\State $q^{\ast} \leftarrow$ \textbf{Quantile}($S^{\ast}_1, ..., S^{\ast}_R,$ ConfidenceLevel)
						\State \textbf{Solve}($\hat S(\beta) =q^{\ast}$) $\rightarrow$ $UpperLimit$
            \State \textbf{return} UpperLimit \Comment{The upper limit of the one-sided FMB CI}
        \EndProcedure
    \end{algorithmic}
\end{Algorithm}
In Algorithm \ref{AlgFMB}, we exploit the monotonicity assumption only in Step 5, where we invert the studentized estimating function $\hat S(\beta)$; see \cite{hu_estimating_2000} for the use of a similar device. 

Indeed, to derive the CI, the bootstrap procedure first provides $(1-\alpha)$-quantile 
estimates of $\hat S(\beta_0)$, say $q^\ast_{1-\alpha}.$
Then, a numerical method (e.g. Newton-Raphson or secant methods) defines a one-sided $(1-\alpha)$-CI for $\beta_0$ as $[\beta_{\min}, \hat{q}_{1-\alpha}]$, where $\beta_{\min}:= \min\mathcal{B}$ and the upper limit $\hat{q}_{1-\alpha}$ solves $\hat S (q_{1-\alpha}) = q^\ast_{1-\alpha}$ in $q_{1-\alpha}.$

For a two-sided equal-tailed CI, we follow the same principles. 
We consider two real numbers $s_1$ and $s_2$ such that $\mathbb{P}[\hat{S} \left( \beta_0 \right) \leq s_1]=\alpha/2$ and $\mathbb{P}[ \hat{S} \left( \beta_0 \right) > s_2 ]=\alpha/2$. From FMB, we obtain the approximation\footnote{As customary in the bootstrap literature, $\mathbb{P}^{\ast}[X^{\ast} \leq x ]$ denotes the empirical c.d.f.\ of any variable $X^{\ast}$ generated by the bootstrap scheme (here $S^\ast$). We give the general definition of the bootstrap probability measure $\mathbb{P}^{\ast}$ in \eqref{boot_distrib}, Section \ref{Theory}.} $\mathbb{P}^{\ast}[s_1 < S^{\ast} \leq s_2 ] = \mathbb{P}[s_1 < \hat{S} ( \beta_0 ) \leq s_2 ] + R_T$, where $R_T$ is an asymptotically  negligible remainder term. Hence, we can compute $s_1$ and $s_2$ such that $\mathbb{P}^{\ast}[s_1 < S^{\ast} \leq s_2 ] = 1 - \alpha$. Then, the CI for $\beta_0$ is $\mathcal{C}_{1-\alpha} := (c_1,c_2]$, with $c_1 := \hat{S}^{-1}\left(s_1\right)$ and $c_2 := \hat{S}^{-1}\left(s_2\right)$, ensuring that $\mathbb{P}\left[c_1 < \beta_0 \leq c_2 \right] = 1 - \alpha + R_T$. Under Assumptions \ref{a1}---\ref{EE5} (Section \ref{Theory}), we can get $R_T = o_p\left(T^{-1/2}\right),$ given a suitable choice of kernel $k$ and bandwidth $B_T$ (see Theorem \ref{higher_order} and discussion in Section \ref{Theory}). It implies that $\mathcal{C}_{1-\alpha}$ is correct up to a higher order.

From the studentization in $\hat S(\beta)$, the CI limits $c_1$ and $c_2$ derived in Step 5 remain invariant to monotonic transformation of the parameter. This property is crucial for the bootstrap CI (\cite{diciccio_bootstrap_1996}), ensuring stability of FMB across varying parameter scale. 

\vspace{1cm}
\textbf{Example 1 [cont'd].}
\textit{ Let us see how the steps in Algorithm 1 specialize for the $AR(1).$ In Step 1, we have $g_{T,t}$ as in \eqref{gTtAR} and $\bar { g }_T ( \beta ) = T ^ { - 1 } \sum _ { t = 1 } ^ { T } {B_{T}}^{-1/2}\sum_{s = t-T}^{t-1} k({s}/{B_{T}}) (Y_{t-s}-\beta Y_{t-s-1})Y_{t-s-1}$. For Step 2, the estimator is available in closed form:
$$
\hat{\beta}=\left( \displaystyle \sum_{t=1}^{T} \displaystyle\sum_{s = t-T}^{t-1} k\left(\frac{s}{B_{T}}\right) Y_{t-s}Y_{t-s-1}\right) \left(\displaystyle \sum_{t=1}^{T} \displaystyle\sum_{s = t-T}^{t-1} k\left(\frac{s}{B_{T}}\right)Y_{t-s-1}^2 \right)^{-1}.
$$
For Step 3 and Step 4, we define $S^{\ast}$ using \eqref{Eq S_psi_ast}, and we use i.i.d.\ resampling of $\{g_{T,t}(\hat{\beta})\}_{t=1}^{T}$. Since $\hat S(\beta)$ is strictly decreasing, we switch the sign of $S^{\ast}$ and $\hat S(\beta)$ and proceed as in Step 5 of Algorithm 1 to build a one-sided CI for $\beta_0$. In this particular case of linear models, we can rewrite $\hat{S}(\beta_0)$ as $T^{1/2}(\hat{\beta}-\beta_0)/\hat{\varsigma},$ where $\hat{\varsigma} = \hat{W}^{-1}\hat{\sigma}$ and $\hat{W} := T^{-1}\sum_{t=1}^{T} \partial g_{T,t}(\hat{\beta})/ \partial \beta.$ Similarly, we can rewrite the bootstrap counterpart $S^{\ast}$ as $T^{1/2}(\beta^{\ast}-\hat{\beta})/\varsigma^{\ast},$ where $\beta^{\ast}$ is a bootstrap estimate, $\varsigma^{\ast} = W^{\ast-1}\sigma^{\ast }$ and $W^{\ast} := T^{-1}\sum_{t=1}^{T} \partial g_{T,t}^{\ast}(\hat{\beta})/ \partial \beta.$ Then, the FMB CI is equivalent to $[\hat{\beta}-q^{\ast}_{1-\alpha/2}\hat{\sigma} , \hat{\beta}-q^{\ast}_{\alpha/2}\hat{\sigma}],$
 where $q^{\ast}$ are quantiles of $S^{\ast}.$ In this representation, the FMB CI is similar to a percentile$-t$ bootstrap CI, up to our use of $\hat{\beta}$ instead of $\beta^{\ast}$ in the definition of $\varsigma^{\ast},$ avoiding to compute $\beta^{\ast}$, and in the multivariate case, to invert a matrix, for each bootstrap sample. This modification does not impact higher-order correctness, as shown in Section \ref{Theory}.}\\

\begin{remark}[Step 5, Lines 16-17]\label{nomono}

When the function $\hat S(\beta)$ is not monotonic, we can slightly modify the procedure if we want to obtain a simply connected CI, as opposed to a union of intervals. To this end, let us define $\hat{Q}(\beta) := \hat S(\beta)^2.$ As an alternative statistic, we take the third-order Taylor expansion of $\hat{Q}(\beta)$ around the root-$T$ consistent estimator $\hat{\beta}.$ Namely, we define $\tilde{Q}(\beta) := (\partial^2 \hat{Q}(\hat{\beta})/\partial \beta^2)(\beta-\hat{\beta})^2/2 + (\partial^3 \hat{Q}(\hat{\beta})/\partial \beta^3)(\beta-\hat{\beta})^3/6,$ the first two terms being zero.  If we make use of $\tilde{Q}(\beta)$ in a neighborhood $\{ \check{\beta} \in \mathcal{B} : \check{\beta} = \beta_0 + \bar{\delta} T^{-1/2} \}$, for  $\bar{\delta} \in \mathbb{R}$  (\cite{newey_notitle_1994}), we can show that $\tilde{Q}(\beta) = \hat{Q}(\beta) + O_p(T^{-1})$ in that neighborhood. 
Hence, FMB allows us to approximate $\mathbb{P}[\tilde{Q}( \beta_0 ) \leq q^{\ast}_{1-\alpha} ]$ by $\mathbb{P}^{\ast}[S^{\ast 2} \leq q^{\ast}_{1-\alpha} ]$ with higher-order accuracy, as shown in Corollary \ref{HOtildeQ} (Section \ref{Theory}). Then, we compute
$\mathcal{C}_{1-\alpha} := \left\{ \beta \in \mathcal{B} : \tilde{Q}\left( \beta \right) \leq q^{\ast}_{1-\alpha} \right\}$ to get the desired higher-order correct $(1-\alpha)$-CI. From Section 9.1 in \cite{newey_notitle_1994}, it should be clear that $\bar \delta$ is not a tuning parameter to be chosen to apply FMB.

This modified FMB CI is simply connected with high probability when $T$ is large enough. Indeed, we can show that $\tilde{q}=\frac{4T}{27\sigma^2}\left[\frac{\partial\bar{g}_T(\hat{\beta})^2}{\partial \beta}/\frac{\partial^2\bar{g}_T(\hat{\beta})}{\partial \beta^2}\right]^2$ is the highest value such that the sublevel set $\left\{ \beta \in \mathcal{B} : \tilde{Q}\left( \beta \right) \leq \tilde{q} \right\}$ is still simply connected. It corresponds to the local maximum of a cubic polynomial (we provide a graphical illustration in Figure \ref{nomonoC}, Supplementary Material SM.9). Thus, the range of confidence level from which we can draw a simply connected set increases proportionally to the sample size $T.$ In practice, we recommend to try using the $\tilde{Q}$ approximation to guarantee the second-order correctness of the CI and connected CI with high probability. If this approach yields a disconnected CI because of a too small sample size $T$, the user can truncate the Taylor approximation at the quadratic term, which gives a simply connected CI w.p.1.\ This quadratic approximation does not guarantee higher-order correctness, but is prone to work better than the Gaussian approximation in practice. To summarise, we face three possibilities. If we use $\hat{Q}$, we always get higher-order correctness, but not necessarily connected CI when monotonicity is not satisfied. If we use the cubic approximation $\tilde{Q}$, we get higher-order correctness and connected CI with high probability. If we use a quadratic truncation, while being asymptotically correct, higher-order correctness might be lost, but we ensure connected CI.
 
Therefore, we conclude that, even if the monotonicity condition is violated (or is not easy to check), we can choose a convenient statistic based on a cubic approximation and preserving the asymptotic properties of FMB. We point out that the proposed derivation of the CI only involves the (potentially numerical) computation of $\hat{Q}'$s derivatives, whose evaluation is required only at the single point $\hat{\beta}$. As the shape of the CI is fully determined by these derivatives, it simplifies and speeds up the implementation of Step 5.
\end{remark}

Some further remarks on the other steps Algorithm \ref{AlgFMB} are in order. First, Step 1 --- Step 3 (Lines 2-13) hinge on bootstrapping the moment indicator evaluated at $\hat\beta$. It justifies the adjective ``fast'' in the name of our resampling scheme, and bears some similarities to the already existing  fast bootstrap  (henceforth FB) methods; see \cite{shao_jackknife_1995}, \cite{davidson_bootstrap_1999}, \cite{andrews_higher-order_2002}, \cite{salibian-barrera_bootstrapping_2002}, \cite{goncalves_maximum_2004}, \cite{salibian-barrera_principal_2006}, \cite{salibian-barrera_fast_2008}, \cite{camponovo_predictability_2013}, \cite{armstrong_fast_2014}, \cite{goncalves_bootstrapping_2019}), and to the estimating function bootstrap (\cite{parzen_resampling_1994}, \cite{hu_estimating_2000}). However, FB methods typically rely on a first order von Mises expansion, which approximation error prevents the FB to be higher-order accurate.

Second, Step 3 (Line 12) computes the bootstrap statistic $S^{\ast},$ where the kernel $k$ creates a block of moment indicators evaluated at $\hat\beta$. The block of $g_t$ induced by the kernel is similar to a moving-average, as we emphasize in the name of our resampling scheme. The Moving Block Bootstrap (henceforth MBB) is the state-of-the-art higher-order correct alternative to FMB (\cite{gotze_second-order_1996}, \cite{lahiri_edgeworth_1996}). For MBB, the blocks are defined at the level of the observations, whereas in our case the convolution is applied to the moment indicators. 

In the same family of groupwise resampling schemes, FMB is even more reminiscent of the Tapered Block Bootstrap of \cite{paparoditis_tapered_2001} (hereafter TBB), in the sense that we can view their tapered block as our moving-average kernel. The main difference is that our kernel has unbounded support, in contradistinction with their block tapering window. It gives FMB an advantage in the studentization: it allows us to use the Quadratic Spectral (QS) kernel, which is optimal in terms of asymptotic mean squared error according to \cite{andrews_heteroscedasticity_1991}. \cite{parente_kernel_2018b} have already pointed out such an advantage for a KBB variance estimator. Yet, the TBB and the KBB approach of \cite{parente_generalised_2018a} both require $R$ bootstrap estimations. Thus, neither the TBB nor the KBB is fast and there is no result on their potential higher-order correctness. The higher-order correctness of the FMB approximation to the distribution of $\hat{S}$ comes from jointly considering two ingredients: the smoothing of moment indicators and the studentization.\footnote{The higher-order correctness of the FMB CI comes from the higher-order correctness of the latter FMB distribution and from the inversion step (Step 5 with the monotonicity condition or the cubic approximation of Remark \ref{nomono}).} Taken in isolation, each ingredient does not allow to show higher-order correctness of the FMB CI.

To summarize, we itemize in Table \ref{RM} the main features of the discussed bootstrap schemes. We only list methodologies that are specifically designed for dependent data.

\begin{table}[h]
\caption{Properties of related bootstrap schemes for dependent data.}
\begin{center}
\begin{tabular}{|l||*{2}{c|}}\hline
\backslashbox{Fast}{HOC}
&\makebox[3em]{Yes}&\makebox[4.5em]{No}\\\hline\hline
\makebox[3em]{Yes} &\makebox[3em]{FMB}&\makebox[4.5em]{FB}\\\hline
\makebox[3em]{No} &\makebox[3em]{MBB}&\makebox[4.5em]{TBB, KBB}\\\hline
\end{tabular}
\label{RM}
\caption*{

We distinguish the Fast Moving-average Bootstrap (FMB), the Fast Bootstrap (FB), the Moving Block Bootstrap (MBB), the Tapered Block Bootstrap (TBB), and the Kernel Block Bootstrap (KBB) with respect to two features, namely computational speed (Fast) and higher-order correctness (HOC).}
\end{center}
\end{table}

\subsection{Over-identified case with multivariate parameter}\label{MParam}

In this section, we explain how FMB can yield higher-order correct inference on a multivariate parameter. In Subsection \ref{Sec: GMM}, we consider the Generalized Method of Moments. Then, we extend the setting to Generalized Empirical Likelihood Estimation in Subsection \ref{Sec: GEL}. The asymptotic refinements (Section \ref{Theory}) also hold for the standard GMM case and are not tied to the use of GEL.

\subsubsection{Generalized Method of Moments} \label{Sec: GMM}

Assume we have to conduct inference on the multivariate parameter $\beta \in \mathcal{B} \subset \mathbb{R}^p$, where $\mathcal{B}$ is compact. We are given a random sample of $\mathbb{X}_t = \left\{ X_{t_1},...,X_{t_v} \right\}$ observed at $t=1,...,T,$ and we define a set of moment conditions $g: \mathbb{R}^{d v} \times \mathcal{B} \rightarrow \mathbb{R}^r$, with $r \geq p$, such that $\mathbb{E}\left[g(\mathbb{X}_t,\beta_0)\right] = 0$. To handle the serial dependence, we define the smoothed moment indicator $g_{T,t}\left(\beta\right)$ as in the univariate case
\begin{align*}
g_{T,t}\left(\beta\right) := {B_{T}}^{-1/2}\displaystyle\sum_{s = t-T}^{t-1} k\left(\frac{s}{B_{T}}\right) g_{t-s}\left(\beta\right),
\end{align*}
\noindent and $\bar{g}_T(\beta) := T^{-1}\sum_{t=1}^{T}g_{T,t}(\beta).$ Estimating $\beta_0$ via Generalized Method of Moments  (\cite{hansen_large_1982}) is the most popular approach in econometrics. In the next subsection, we discuss alternative estimators.

Step 1 --- Step 3 of the FMB Algorithm \ref{AlgFMB} remain conceptually 
unchanged. As far as the bootstrap statistic is concerned, the principles of Step 4 and Step 5 stay the same as in Algorithm \ref{AlgFMB}, 
the main change being that the asymptotically pivotal statistic becomes:
\begin{align}
\hat Q  := T \bar{g}_{T} \left( \beta_0 \right) ^\intercal \hat{\Omega}^{-1} \bar{g}_{T} \left( \beta_0 \right),\label{Rao_type}
\end{align}
where $\hat{\Omega} = \kappa_1^2(\kappa_2 T)^{-1}\sum_{t = 1}^{T} \{ g_{T,t}( \hat \beta ) - \bar{g}_T( \hat \beta ) \} \{ g_{T,t}( \hat \beta ) - \bar{g}_T( \hat \beta )\}^{\intercal}$ is a consistent estimator of the long-run covariance matrix of $T^{1/2}\bar{g}_{T} \left( \beta_0 \right)$, of rank $\nu = r$;\footnote{If the rank is lower than $r$, the covariance matrix is not invertible anymore and we resort to the generalized inverse, adapting the degrees of freedom of the $\mathcal{X}^2$ distribution accordingly (\cite{moore_generalized_1977}).} see Section \ref{sectionHAC}. Standard results guarantee that $\hat Q$ is asymptotically $\mathcal{X}^2_{\nu}$. As in the univariate case, the statistic of interest is a particular value of a function, here $\hat{Q}: \mathbb{R}^{d v} \times \mathcal{B} \rightarrow \mathbb{R},$
\begin{equation}
\hat{Q}(\beta) :=  T \bar{g}_{T} \left( \beta \right) ^\intercal \hat{\Omega}^{-1} \bar{g}_{T} \left( \beta \right). \label{Eq. hatQ}
\end{equation}

\noindent We define the GMM estimator as $\hat{\beta} = \argmin_{\beta \in \mathcal{B}} \hat{Q}(\beta).$ Similarly to the argument of Remark \ref{nomono}, we also define the cubic approximation centered on the root-$T$ consistent estimator $\hat{\beta}:$
\begin{align}
\tilde{Q}(\beta) := \hat{Q}(\hat{\beta}) + (\beta-\hat{\beta})^{\intercal} \hat{H} (\beta-\hat{\beta}) /2 + \left( (\beta-\hat{\beta}) \otimes (\beta-\hat{\beta}) \right)^{\intercal} \frac{\partial \vec2 (\hat{H})}{\partial \beta^{\intercal}} (\beta-\hat{\beta}) /6,
\label{tildeQStat}
\end{align}
\noindent where the matrix $\hat{H} := \partial^2 \hat{Q}(\hat{\beta}) / \partial \beta^{\intercal} \partial \beta$. It allows us to build simply connected level sets, yielding higher-order correct Confidence Region (henceforth CR), as shown in Corollary \ref{HOtildeQ} (Section \ref{Theory}).

The bootstrap version of $\hat Q$ is 
\begin{align}
Q^{\ast} := T^{-1} \sum_{t=1}^{T} \left\{ g^{\ast}_{T,t} - \bar{g}_T\left( \hat\beta \right) \right\} ^\intercal \hat{\Omega}^{\ast-1} \sum_{t=1}^{T} \left\{ g^{\ast}_{T,t} - \bar{g}_T\left( \hat\beta \right) \right\}\label{Rao_type_ast},
\end{align}
where 
$
\hat{\Omega}^{\ast} := {T}^{-1}\sum_{t = 1}^{T} \left\{ g^{\ast}_{T,t} - \bar{g}_T\left( \hat\beta \right) \right\} \left\{ g^{\ast}_{T,t} -\bar{g}_T\left( \hat\beta \right)\right\}^{\intercal},
$
with the asterisk denoting the same i.i.d.\ resampling scheme as in Algorithm \ref{AlgFMB}. Now we are ready to state the algorithm of our FMB in the over-identified case, made of five steps (lines 2-4, 5, 6-13, 14-15, 16-17).
\begin{Algorithm}\label{Alg2FMB}
\textnormal{\vspace{0.01cm}}
    \begin{algorithmic}[1] 
            \Procedure{FMB}{$X_1, ..., X_T,$  $g,$ $\alpha$} \Comment{The FMB $(1-\alpha)$-CR for $\beta_0 \in \mathcal{B}$ s.t. $\mathbb { E } \left[ g \left( \mathbb{X} _ { t } , \beta _ { 0 } \right) \right] = 0.$}
            \For{$t = 1, ..., T$}
						    \State $g_{T,t}\left(\beta\right) \leftarrow {B_{T}}^{-1/2}\sum_{s = t-T}^{t-1} k\left(s/B_{T}\right) g_{t-s}(\beta)$
						\EndFor\label{Alg2FMBconvol}
						\State \textbf{Argmin}$_{\beta \in \mathcal{B}} \hat{Q}(\beta)$ $\rightarrow$ $\hat{\beta}$ \Comment{Using the function $\hat{Q}(\beta)$ as in \eqref{Eq. hatQ}.}
						\For{$r = 1, ..., R$}
						
						    \For{$t = 1, ..., T$}
						         \State $g_{T,t}^{\ast} \leftarrow$ \textbf{Draw}($g_{T,1}(\hat{\beta}), ..., g_{T,T}(\hat{\beta})$)
						    \EndFor\label{Alg2FMBSample}
						    \State $\bar{g}_{T,r}^\ast \leftarrow T^{-1} \sum_{t=1}^{T} \left\{ g^{\ast}_{T,t} - \bar{g}_T\left( \hat\beta \right) \right\}$
						    \State $\hat{\Omega}_r^{\ast} \leftarrow {T}^{-1}\sum_{t = 1}^{T} \left\{ g^{\ast}_{T,t} - \bar{g}_T\left( \hat\beta \right) \right\} \left\{ g^{\ast}_{T,t} -\bar{g}_T\left( \hat\beta \right)\right\}^{\intercal}$
						    \State $Q^{\ast}_r := T \bar{g}_{T,r}^{\ast \intercal} \hat{\Omega}_r^{\ast-1} \bar{g}_{T,r}^\ast$
						\EndFor\label{Alg2FMBBoot}
						\State ConfidenceLevel $\leftarrow$ $1-\alpha$
						\State $q^{\ast} \leftarrow$ \textbf{Quantile}($Q^{\ast}_1, ..., Q^{\ast}_R,$ ConfidenceLevel)
            \State $\mathcal{C}$ $\leftarrow$ $\left\{ \beta \in \mathcal{B} : \hat{Q}\left( \beta \right) \leq q^{\ast} \right\}$
						\State \textbf{return} $\mathcal{C}$
        \EndProcedure
    \end{algorithmic}
\end{Algorithm}
A few remarks are in order. Step 3 uses \eqref{Rao_type_ast}, where we recenter the bootstrap statistic. Indeed, the bootstrap expectation $\mathbb{E}^{\ast}\left[g^{\ast}_{T,t}\right] = \bar{g}_T(\hat{\beta}) \neq 0$ in the over-identified case. Thus, we subtract its expectation from $g_{T,t}^{\ast}$ to recenter the bootstrap variable. This operation is crucial to achieve higher-order accurate CR in Step 5 of Algorithm \ref{Alg2FMB} (see e.g.\ \cite{hall_bootstrap_1996}).

Moreover, in contradistinction with the already existing FB methods, Step 3---4 mimic the variability of the covariance estimator $\hat{\Omega}$ (in \eqref{Rao_type}) to achieve higher-order refinements. To this end, we use $\hat{\Omega}^{\ast}$ instead of $\hat{\Omega}$ in the definition of $Q^{\ast}$ (in \eqref{Rao_type_ast}), such that we randomize the bootstrap covariance estimator across the different bootstrap samples, and do not keep it fixed at $\hat{\Omega}$. Similar comment applies to Algorithm \ref{AlgFMB}.

Finally, to define the CR for $\beta_0$, we proceed similarly to Step 5 of Algorithm \ref{AlgFMB}. We set $q^{\ast}_{1-\alpha}$ such that $\mathbb{P}^{\ast}[ Q^{\ast} \leq q^{\ast}_{1-\alpha} ] = 1-\alpha$ and compute the CR as the subset $\mathcal{C}_{1-\alpha} := \left\{ \beta \in \mathcal{B} : \hat{Q}\left( \beta \right) \leq q^{\ast}_{1-\alpha} \right\}$. Thus, we get $\mathbb{P}[ \beta_0 \in \mathcal{C}_{1-\alpha} ] = \mathbb{P}[ \hat{Q} ( \beta_0 ) \leq q^{\ast}_{1-\alpha}] = \mathbb{P}^{\ast} [ Q^{\ast} \leq q^{\ast}_{1-\alpha}] + R_T = 1-\alpha + R_T$. Under Assumptions \ref{a1}---\ref{EE5} (Section \ref{Theory}), we show in Theorem \ref{higher_order} that the remainder $R_T$ can be at most of order $o_p\left(T^{-1/2}\right),$ given a suitable choice of kernel $k$ and bandwidth $B_T$, which implies that $\mathcal{C}_{1-\alpha}$ is correct up to a higher order.

\begin{remark}[Step 5, Lines 16-17]\label{nomono2} 

If the higher-order correctness of FMB is guaranteed independently of the CR shape, they are generally not elliptical and they might fail to be simply connected (they can come in several pieces or contain holes). Although this irregularity is customary for small to moderate sample sizes, it can still make the results difficult to interpret. The monotonicity condition aforementioned is one of the possible ways out. As a more general solution, we propose to use the cubic approximation $\tilde{Q}$ (as in \eqref{tildeQStat}). Similarly to Remark \ref{nomono}, we have $\tilde{Q}(\beta) = \hat{Q}(\beta) + O_p(T^{-1})$ in a neighborhood $\{ \check{\beta}  \in \mathcal{B} : \check{\beta} = \beta_0 + \bar{\delta} T^{-1/2}\},$ for $ \bar{\delta} \in \mathbb{R}^p.$  Thus, defining the modified FMB CR as $\mathcal{C}_{1-\alpha} := \left\{ \beta \in \mathcal{B} : \tilde{Q}\left( \beta \right) \leq q^{\ast}_{1-\alpha} \right\}$ preserves higher-order correctness, as shown in Corollary \ref{HOtildeQ}. The resulting CR is not simply connected for all sample sizes, but we can show that the range of confidence level $(1-\alpha)$ leading to a simply connected CR is proportional to the sample size $T.$ It is not an asymptotic property and the desired CR can already be simply connected for a small sample size, depending on the second derivative of the moment indicator $g.$ If the sample size is too small for this modified CR to be simply connected and the user thoroughly needs this property, we recommend to truncate the $\tilde{Q}$ approximation at the second term. The resulting CR is always simply connected and elliptical, but, while being asymptotically correct, we cannot guarantee its higher-order properties.

\end{remark}

\subsubsection{Generalized Empirical Likelihood Estimation} \label{Sec: GEL}

FMB is not tied to a particular estimation method. In order to show its wide applicability, we consider a general estimation method, which includes (among others) a version of GMM. The Continuously Updating Estimator (CUE) (\cite{hansen_finite-sample_1996}), Empirical Likelihood (EL) (\cite{qin_empirical_1994}, \cite{imbens_one-step_1996}, \cite{kitamura_empirical_1997}), and the Exponential Tilting (ET) (\cite{kitamura_information-theoretic_1997}, \cite{imbens_information_1998}) are all asymptotically equivalent to the (2S)GMM, but they tend to be less biased for small to moderate sample sizes (see \cite{altonji_small-sample_1996} for a Monte Carlo exploration, and \cite{newey_higher_2004}, \cite{anatolyev_gmm_2005} for theoretical insights). Putting EL, ET, and CUE  under the same umbrella, \cite{smith_gel_2011} introduces the Generalized Empirical Likelihood (GEL) criterion for time series data. We briefly describe it before extending our FMB to this general setting.

Let $\rho \left( \nu \right)$ be a concave function on an open interval $\mathcal{V} \in \mathbb{R}$ containing $0$.
Writing $\rho_{\iota}(\nu) := \partial^{\iota} \rho (\nu) / \partial \nu^{\iota}$ 
and $\rho_\iota = \rho_{\iota}(0)$ for $\iota=0,1,$ the function
$\rho \left( \nu \right)$ is standardized such that $\rho_1 = -1$. 
Defining a vector of auxiliary parameters $\lambda \in \Lambda_ { T } ( \beta )$ with $\Lambda_ { T } ( \beta ) := \left\{ \lambda \in 
\mathbb{R} ^ { r } : \kappa B_{T}^{-1/2} \lambda ^ { \intercal } g _ { T,t } ( \beta ) \in \mathcal{V} \right\},$ and $\kappa := \kappa_1 / \kappa_2$, \cite{smith_gel_2011} defines the GEL criterion as:
\begin{align}
\hat{P}(\beta,\lambda) = T^{-1} \sum_{t=1}^{T} \left[ \rho \left( \kappa B_{T}^{-1/2} \lambda ^\intercal g_{T,t} \left( \beta \right) \right) - \rho_0 \right].\label{GEL_criterion}
\end{align}
\noindent To derive an estimator of $\beta$, we first optimize criterion \eqref{GEL_criterion} w.r.t.\ $\lambda$ for a given $\beta$, so that $\lambda \left( \beta \right) = \argsup_{\lambda \in \Lambda_ { T } ( \beta )} \hat{P}(\beta,\lambda)$. Then, we define $\hat{\beta}$ as the solution to $\argmin_{\beta \in \mathcal{B}} \hat{P}(\beta,\lambda\left( \beta \right))$.

Similarly to \cite{khundi_edgeworth_2012} and \cite{lee_asymptotic_2016}, we are going to use the first order condition of the GEL criterion as a just-identified representation of the estimation problem to explain how to build the FMB statistic. Differentiating \eqref{GEL_criterion} w.r.t.\ to $\lambda$ and $\beta$, we obtain:
\begin{align}
&T^{-1}\displaystyle\sum_{t=1}^{T} \rho_{1} \left( \kappa B_{T}^{-1/2} \lambda \left( \beta \right) ^\intercal g_{T,t}\left( \beta \right) \right) B_{T}^{-1/2} {g}_{T,t}\left( \beta \right) = 0,\label{GELFOC1}
\\
&T^{-1}\displaystyle\sum_{t=1}^{T} \rho_{1} \left( \kappa B_{T}^{-1/2} \lambda \left( \beta \right) ^\intercal g_{T,t} \left( \beta \right) \right) B_{T}^{-1/2}\frac{ \partial g_{T,t} \left( \beta \right)}{\partial \beta }^\intercal \lambda \left( \beta \right) = 0.\label{GELFOC2}
\end{align}
\noindent We can see from \eqref{GELFOC1} that $\rho_{1} \left( \kappa B_{T}^{-1/2} \lambda \left( \beta \right) ^\intercal g_{T,t}\left( \beta \right) \right)$ gives weights to the observations such that the moment conditions in $g$ are always enforced in a given sample. GEL estimators are equivalent to some minimum discrepancy estimators based on the power-divergence family (see \cite{cressie_multinomial_1984}), where the auxiliary vector parameter $\lambda$ corresponds to the Lagrange multiplier enforcing this empirical moment condition. Thus, if the original moment conditions in $g$ are correctly specified, the true (long-run) Lagrange multiplier is zero ($\lambda_0 = 0$). Applying our FMB to this setting only requires to define a quadratic statistic from the GEL first order conditions (\eqref{GELFOC1} and \eqref{GELFOC2}) evaluated at the true value of the parameters. Yet, replacing $\lambda = \lambda_0 = 0$ and $\beta = \beta_0$ in \eqref{GELFOC1} and \eqref{GELFOC2} boils down to the original $\bar{g}_T(\beta_0).$ Thus, the natural extension of FMB for GEL estimators requires to take the asymptotically pivotal statistic $\hat{Q}(\beta_0)$ as in the GMM case \eqref{Rao_type}. As a consequence, FMB in the GEL setting is exactly the same as in the GMM setting (see Algorithm \ref{Alg2FMB}), up to the initial estimator $\hat{\beta}.$  It is quite  intuitive since, in absence of misspecification, the first order conditions of the GEL criterion convey the same information on $\beta_0$ as the moment condition $g$.

\section{Theory}\label{Theory}

In the next theorems, we state that FMB CI and CR are higher-order correct. By construction, the higher-order correctness of the FMB CI and CR entirely hinges on our bootstrap approximation of the distribution for the test statistics $\hat{S}(\beta_0)$ (as in \eqref{Eq S_psi}) and $\hat{Q}(\beta_0)$ (as in \eqref{Rao_type}). 

We start by itemizing the assumptions and regularity conditions. In the following, we keep using the shorthand notations $g_t = g_t\left(\beta_0\right)$ and $g_{T,t} = g_{T,t}(\beta_0).$ For any vector $V \in \mathbb{R}^n$, we write $\| V \| = (v_1^2 + ... + v_n^2)^{1/2}$, where $v_j$ is the $j$-th element of $V$. We make use of generic constants $C$, $\delta$ and $\epsilon$, whose value can differ from an expression to another. We define $\left\{g_t\right\}_{t \in \mathbb{Z}}$ on the probability space $\left(\Omega,\mathcal{A},\mathbb{P}\right)$. Let $\left\{ \mathcal{D}_t \right\} _ {t \in \mathbb{Z}}$ be a given sequence of sub-sigma-fields of $\mathcal{A}$, and $\mathcal{D}_{a}^{b} = \sigma\left\langle\left\{\mathcal{D}_{j} : a \leq j \leq b \right\}\right\rangle$. A straightforward example is to take $\mathcal{D}_t:=\sigma\left\langle g_t \right\rangle,$ but it is not always the most efficient choice to check the assumptions below (see \cite{gotze_asymptotic_1983} and \cite{gotze_asymptotic_1994} for practical examples). The higher-order correctness of FMB is subject to the following conditions (\cite{gotze_second-order_1996}, \cite{lahiri_resampling_2010}), which we assume to hold for $\left\{g_t\right\}_{t \in \mathbb{Z}}$:

\begin{asu}\label{a1}
$\mathbb{E}\left[g_{t}(\beta)\right] = 0, t=1,2,...$ only for $\beta = \beta_0$ on the compact parameter space $\mathcal{B}$. Moreover, $\hat{\beta} \overset{a.s.} \rightarrow \beta_0$ and $\|\hat{\beta}-\beta_0\|=O_p(T^{-1/2})$ as $T \rightarrow \infty$.
\end{asu}
\begin{asu}\label{a2}
$\mathbb{E}\left[\left\| g_{t} \right\|^{s+\delta}\right] < \infty \text{ for a positive } s \geq 8,\text{ }t=1,2,...,\text{ and } \delta > 0.$
\end{asu}
\begin{asu}\label{RV_approx}
There exists a constant $\delta > 0$ such that for $t,m = 1,2,...$ and $m > \delta^{-1}$, we can approximate $g_{t}$ by a $\mathcal{D}_{t-m}^{t+m}$-measurable random vector $g_{t,m}^{\ddagger}$, such that $\mathbb{E}\left\| g_t - g_{t,m}^{\ddagger} \right\| \leq \delta^{-1} \exp\left(-\delta m\right)$.
\end{asu}
\begin{asu}\label{Rosenblatt_mix}
There exists a constant $\delta > 0$ such that
$\alpha\left(m\right) = \sup \left\lvert \mathbb{P}\left[A \cap B\right] - \mathbb{P}\left[A\right]\mathbb{P}\left[B\right]\right\rvert \leq \delta^{-1} \exp\left(-\delta m\right)$, for all $t,m = 1,2,... $ and $A \in \mathcal{D}_{-\infty}^{t}, B \in \mathcal{D}_{t+m}^{\infty}$.
\end{asu}
\begin{asu}\label{Condition_Cramer}
There exists a constant $\delta > 0$ such that for all $t,m = 1,2,...$, $\delta^{-1} < m < t$, and all $\tau \in \mathbb{R}^r$ with $\| \tau \| \geq \delta$,
$\mathbb{E}\left[\left\lvert\mathbb{E}\left[\exp\left(i\tau ^{\intercal} \left(g_{t-m} + ... + g_{t+m}\right) \right) \middle| \mathcal{D}_k : k \neq t\right]\right\rvert\right] \leq \exp\left(-\delta\right)$, and
\begin{align}
\displaystyle\liminf_{T \rightarrow \infty} \frac{1}{T} \Var \left[ \displaystyle\sum_{t=1}^{T} g_t \right] > 0.
\end{align}
\end{asu}
\begin{asu}\label{Markov_type}
There exists a constant $\delta > 0$ such that for all $t,m,p = 1,2,...$ and $A \in \mathcal{D}_{t-p}^{t+p}$, \newline
$\mathbb{E}\left[\left\lvert \mathbb{P}\left[A\middle|\mathcal{D}_k: k \neq t\right] - \mathbb{P}\left[A\middle|\mathcal{D}_k: 0 < \left\lvert t-k \right\rvert \leq m + p\right]\right\rvert\right] \leq \delta^{-1} \exp\left(-\delta m\right)$.
\end{asu}
\begin{asu}\label{EE5}
Defining $f(\beta):=\limsup_{T \rightarrow \infty} \sup_{b < \| \tau \| < e^{\delta T}} \lvert T^{-1} \displaystyle\sum_{t=1}^{T} \exp \left( i  \tau^{\intercal} g_{T,t}\left(\beta\right) \right) \rvert,$ there exist constants $b$ and $\delta>0$ such that $f(\beta_0)<1$ and is continuous at $\beta_0$ a.s.
\end{asu}
Assumption \ref{a1} is an identification condition. It is necessary also because we evaluate the moment conditions at $\hat{\beta}$ in the bootstrap samples. Since we are going to prove the validity and higher-order correctness of FMB using  the $(s-2)$-th order Edgeworth expansion for the mean of \cite{gotze_asymptotic_1994}, we require the moments in Assumption \ref{a2} to be defined. Assumption \ref{RV_approx} ensures that the process $\{g_t\}$ is close enough to another process $\{g_{t,m}^{\ddagger}\}$, measurable w.r.t.\ sub-sigma-fields belonging to the sequence $\{\mathcal{D}_t\}$, whose dependence structure is controlled by the mixing condition in Assumption \ref{Rosenblatt_mix}. Assumption \ref{Condition_Cramer} is the conditional Cramér condition of \cite{gotze_asymptotic_1994} for weakly dependent process. Assumption \ref{Markov_type} ensures that we can approximate the probability of $A \in \mathcal{D}_{t-p}^{t+p}$ given $\{ \mathcal{D}_k: k \neq t \}$ with an exponentially increasing accuracy, as the information in $\{ \mathcal{D}_k: 0 < \left\lvert t-k \right\rvert \leq m + p \}$ increases with $m$. We need Assumption \ref{EE5} on the regularity of the bootstrap characteristic function in order for the appropriate Cramér condition to hold for $S^{\ast}$ and $Q^{\ast}$. We refer to \cite{gotze_asymptotic_1983} for a general overview of the processes in agreement with our assumptions. As an example, the OLS moment indicators for the autoregressive parameters of a stationary AR($p$) process $Y_t = \sum_{i=1}^{p} \theta_i Y_{t-i} + e_t = \sum_{j=0}^\infty w_j e_{t-j}$ satisfy Assumptions \ref{a1}---\ref{EE5}, when $e_t \overset{i.i.d.} \sim \mathcal{N}(0,\sigma^2)$ and $\lvert w_j \rvert \leq \delta^{-1} \exp(- \delta j)$ $\forall j \in \mathbb{N},$ $\delta >0$. We verify the assumptions for this example in the Supplementary Material (SM.10). We can check along the same lines that an ACD($v$,0) model, and in particular the ACD(1,0) used in our Monte Carlo simulations, satisfies Assumptions \ref{a1}---\ref{EE5}.

Under the stated assumptions, we prove (see Appendix) the following two theorems, in which we denote by $\mathbb{P}^\ast$ the bootstrap probability measure given the data:
\begin{align}
\mathbb{P}^{\ast}(A) := T^{-1}\sum_{t=1}^{T}\mathbb{I}_{A}\{g_{T,t}(\hat{\beta})\},\label{boot_distrib}
\end{align}
where $A$ is a set and $\mathbb{I}_{A}\{V\} = 1$ if $V \in A$ and 0 otherwise.
\begin{theorem}\label{higher_order}
Under Assumptions \ref{a1}---\ref{EE5}, $\mathbb{E} \left[ \left\| g_t \right\|^{\bar q s + \delta} \right] < \infty$, for $\delta > 0$, $s \geq 8$, and $\bar q \geq 3$, $\log T = o(B_T)$, we have (i) for $S^{\ast}$ as in \eqref{Eq S_psi_ast} and $\hat{S}_{}$ as in \eqref{Eq S_psi}:  
$$\sup_{x \in \mathbb{R}}\left\lvert \mathbb{P}^\ast\left[S^{\ast} \leq x\right] - \mathbb{P}\left[\hat{S}_{} \leq x\right]\right\rvert = o_p\left(T^{-1/2}\right) + O_p\left(B_T^{-q}\right) + O_p\left(B_T/T \right),
$$ 
and (ii) for $Q^{\ast}$ as in \eqref{Rao_type_ast} and $\hat{Q}$ as in \eqref{Rao_type}: $$\sup_{x \in \mathbb{R}^{+}}\left\lvert \mathbb{P}^\ast[ Q^{\ast} \leq x] - \mathbb{P}[\hat{Q} \leq x]\right\rvert = o_p(T^{-1/2}) + O_p\left(B_T^{-q}\right) + O_p\left(B_T/T \right),$$
where $q$ is the Parzen exponent of $k^{\ast}$,  where $k^{\ast}\left( a \right) := \kappa_2^{-1} \int{k\left( b-a \right)k\left( b \right)db}$.
\end{theorem}
It is now apparent that the kernel $k$ impacts the bootstrap accuracy through the variance estimators in $\hat{S}$ and $\hat{Q}.$ As discussed by \cite{parzen_consistent_1957} and \cite{andrews_heteroscedasticity_1991}, the bias of this estimator depends on the smoothness of the kernel $k^{\ast}$ at zero. The kernel $k^{\ast}\left( a \right) := \kappa_2^{-1} \int{k\left( b-a \right)k\left( b \right)db}$ is induced by the self-convolution of the smoothing kernel $k$. This bias is of order $O(B_T^{-q})$, where $q$ is the Parzen exponent of $k^{\ast}$, namely the maximal natural number such that $\lim_{a \rightarrow 0} \left(1-k^{\ast}(a)\right)/ \lvert a \rvert^{q}$ is finite. The bias is minimal for the rectangular kernel. However, as discussed in Section \ref{sectionHAC}, the resulting estimate $\hat{\Omega}$ is not necessarily positive semi-definite. In contrast, the QS kernel has an optimal Parzen exponent $q=2$ over all the kernels giving positive semi-definite estimators. Thus, the covariance matrix estimator with QS kernel has asymptotic MSE of order $O(B_T/T) + O\left( B_T^{-2} \right)$. Consequently, $B_T$ must grow faster than $T^{1/4}$ and slower than $T^{1/2}$, for the bootstrap error to be $o_p\left( T^{-1/2} \right)$.
In Section \ref{method}, we define the FMB CR by $\mathcal{C}_{1-\alpha} = \left\{ \beta \in \mathcal{B} : \hat{Q}\left( \beta \right) \leq {q}^{\ast}_{1-\alpha} \right\}$ (or alternatively using $\tilde{Q}(\beta)$ as in \eqref{tildeQStat}), 
 where ${q}^{\ast}_{1-\alpha}$ is the approximation of $\hat{Q}(\beta_0)$ quantiles by FMB. Thus, $\mathbb{P}[ \beta_0 \in \mathcal{C}_{1-\alpha} ] = \mathbb{P}[ \hat{Q} ( \beta_0 ) \leq {q}^{\ast}_{1-\alpha}] = \mathbb{P}^{\ast} [ Q^{\ast} \leq {q}^{\ast}_{1-\alpha}] + R_T = 1-\alpha + R_T$. Theorem \ref{higher_order} gives the order $R_T = o_p\left(T^{-1/2}\right) + O_p\left(B_T^{-q}\right) + O_p\left(B_T/T \right).$ If we select a kernel $k$ and a bandwidth $B_T$ such that $O_p\left(B_T^{-q}\right) + O_p\left(B_T/T \right) =  o_p\left(T^{-1/2}\right),$ it implies that $\mathcal{C}_{1-\alpha}$ is higher-order correct by construction, as the (first-order) Gaussian approximation is at best of order $O(T^{-1/2}).$
 If $B_T = C T^{\gamma}$, $C > 0,$ we get the conditions $q >1$ and $(2q)^{-1} <\gamma  < 1/2$. Similar arguments apply for Algorithm \ref{AlgFMB}. Part (ii) extends the higher-order correctness of the i.i.d.\ bootstrap of \cite{hu_estimating_2000} in the just-identified multivariate parameter case (see their Remark 10) to the over-identified case with time-dependent data.

When the monotonicity condition discussed in Section \ref{method} (Example \ref{exAR1}) is violated (or is uneasy to check), we may use the alternative FMB CI or CR defined as $\mathcal{C}_{1-\alpha} := \left\{ \beta \in \mathcal{B} : \tilde{Q}\left( \beta \right) \leq q^{\ast}_{1-\alpha} \right\}$ (see Remark \ref{nomono} and \eqref{tildeQStat}), which are simply connected when $T$ is large enough and centered at $\hat{\beta}$. The following corollary (see the Supplementary Material for its proof) states the higher-order correctness of this alternative version of FMB 
CI and CR, based on the $\tilde{Q}$ statistic.
\begin{corollary}\label{HOtildeQ}
Under Assumptions \ref{a1}---\ref{EE5}, $\mathbb{E} \left[ \left\| g_t \right\|^{\bar q s + \delta} \right] < \infty$, for $\delta > 0$, $s \geq 8$, and $\bar q \geq 3$, $\log T = o(B_T)$, we have for $\tilde{Q}$ as in \eqref{tildeQStat} and $Q^{\ast}$ as in \eqref{Rao_type_ast}: $$\sup_{x \in \mathbb{R}^{+}}\left\lvert \mathbb{P}^\ast[ Q^{\ast} \leq x] - \mathbb{P}[\tilde{Q} \leq x]\right\rvert = o_p(T^{-1/2}) + O_p\left(B_T^{-q}\right) + O_p\left(B_T/T \right),$$
where $q$ is the Parzen exponent of $k^{\ast}$.
\end{corollary}
As a consequence, we have $\mathbb{P}[ \beta_0 \in \mathcal{C}_{1-\alpha} ] = \mathbb{P}[ \tilde{Q} ( \beta_0 ) \leq {q}^{\ast}_{1-\alpha}] = \mathbb{P}^{\ast} [ Q^{\ast} \leq {q}^{\ast}_{1-\alpha}] + R_T = 1-\alpha + R_T,$ with $R_T = o_p\left(T^{-1/2}\right) + O_p\left(B_T^{-q}\right) + O_p\left(B_T/T \right).$ Thus, similarly to the CI and CR based on $\hat{S}$ and $\hat{Q},$ the CI and CR based on $\tilde{Q}$ are higher-order correct if we select a kernel $k$ such that $O_p\left(B_T^{-q}\right) + O_p\left(B_T/T \right) =  o_p\left(T^{-1/2}\right)$.

\section{Implementation aspects}\label{implement}

\subsection{Consistent covariance matrix estimation} \label{sectionHAC}

In this section, we give the necessary details on the appropriate way to estimate the long-run variance in the statistics $\hat{S}$ (as in \eqref{Eq S_psi}) and $\hat{Q}$ (as in \eqref{Rao_type}). We treat the general case of $\hat{\Omega}$ (as in \eqref{Rao_type}), from which the univariate case can be easily deduced. 
Following \cite{andrews_heteroscedasticity_1991} and \cite{smith_gel_2011}, we can obtain several estimators $\hat{\Omega}$ using different types of kernels. Let us give some examples of kernels useful in the implementation of FMB and their respective properties.

\begin{example}[Truncated kernel] The so-called truncated kernel is $k(x)=1$, if $|x| \leq 1$, and $0$, if $|x| > 1$. Considering $m_T$ such that $B_T=(2m_T+1)/2$, we write $g_{T,t}(\beta)=2(2m_T+1)^{-1} \sum_{s=\max{[t-T,-m_T]}}^{\min{[t-1,m_T]}} g_{t-s}(\beta).$ The corresponding long-run variance estimator follows directly from the definition of $\hat{\Omega}$ and has minimal asymptotic mean squared error (see \cite{andrews_heteroscedasticity_1991}), but does not guarantee the resulting variance estimator to be positive semi-definite.
The spectral window generator of the truncated kernel is its Fourier transform $K(\lambda)=\pi^{-1}[\sin(\lambda)/\lambda]$.
The corresponding induced kernel is the Bartlett kernel $k^{\ast}(x)=1-|x/2|$ for $|x| \leq 2$, $0$ for $|x| > 2$, as its spectral window  generator is $K^{\ast}(\lambda)=\pi^{-1}[\sin(\lambda)/\lambda]^2.$ According to \cite{andrews_heteroscedasticity_1991}, the corresponding optimal bandwidth parameter is $m^{opt}_T = O(T^{1/3}),$ and the standardizing constants are $\kappa_1 = 2$ and $\kappa_2 = 2.$
\end{example}

\begin{example}[Quadratic Spectral kernel] \label{QS kernel} Among the available kernels ensuring the long-run variance estimator to be positive semi-definite, \cite{andrews_heteroscedasticity_1991} points out the optimal QS kernel $k_{QS}^{\ast}$, as well as the respective optimal bandwidth $B_T^{opt} = O\left( T^{1/5} \right)$. From the relationship $K^{\ast}\left(\lambda\right)= \left(2\pi / \kappa_2\right) \left\lvert K\left( \lambda \right) \right\rvert^{2}$ and the inverse Fourier transform, \cite{smith_gel_2011} identifies the kernel
\begin{align*}
&k_{J}\left( x \right) := \begin{cases}
&\frac{1}{x}J_1 \left( \frac{6 \pi x}{5} \right)(\frac{5 \pi}{8})^{1/2}\text{ if }x \neq 0,
\\
&\frac{3\pi}{5}(\frac{5 \pi}{8})^{1/2}\text{ if }x = 0,
\end{cases} \nonumber
\\
&J_{\nu}\left( z \right) := \frac{z^{\nu}}{2^{\nu}} \displaystyle \sum_{j = 0}^{\infty} \left( -1\right)^{j} \frac{z^{2j}}{2^{2j}j ! \Gamma \left( \nu + j + 1 \right)},\nonumber
\end{align*}
\noindent inducing the QS kernel by self-convolution: $k^{\ast}_{QS} ( a ) = \left( 1 / \kappa _ { 2 } \right) \int k_{J} ( b - a ) k_{J} ( b ) d b$. The standardizing constants are $\kappa_1 = (5\pi/2)^{1/2}$ and $\kappa_2 = 2\pi.$
\end{example}

Thereby, we preferably use $k_{J}\left( x \right)$ of Example \ref{QS kernel} in \eqref{convol} to get the estimator $\hat{\Omega}$. Indeed, both from theory and simulations, the QS kernel is the optimal induced kernel in terms of asymptotic mean squared error (\cite{andrews_heteroscedasticity_1991}), among all kernels giving positive semi-definite long-run variance estimators. The optimal bandwidth $B_T = O (T^{1/5})$ minimising the mean-squared error of the variance estimator with $k_J(x)$ (see Example 5) does not satisfy the condition $(1-\epsilon)/q < \gamma < \epsilon$ for any $\epsilon \in (0,1/2]$ (see Section 3) since $q = 2$ for the QS kernel. Then, we can take $\gamma = 1/3$, so that we match the condition for $q = 2$.  \cite{wilhelm_optimal_2015}  has already exhibited a discrepancy between the optimal choice for the HAC variance estimator and the optimal choice for a GMM point estimator. In his case, the optimal bandwidth for the point estimate is of the same order as the one minimizing the mean-squared error of the nonparametric plugin estimate, while  the constants of proportionality are significantly different. In our case, the order is even different if we want to achieve higher-order correctness for FMB.

Alternatively, we may use the flat-top kernel version of the QS kernel, see \cite{politis_higher-order_2011}, to get an even faster rate of convergence for the estimated long-run variance. Unfortunately, self-convolution of a kernel $k$ cannot induce a flat-top kernel $k^{\ast}$. Indeed, we know that it cannot be the case that $U = X + Y$, where the random variable $U$ is uniformly distributed on $[0, 1]$ (the flat-top part) and the random variables $X$ and $Y$ are independent and identically distributed (see  Exercise 4.14.20 and its proof by contradiction in \cite{grimmett_thousand_2001}). As a consequence, if we want to benefit from the smaller bias of the flat-top kernel, we should use a different kernel for the original statistic and the bootstrap one. A potential modification of FMB is to decouple the kernel $k^{\ast}$ used in the variance estimator of the original statistic, say a flat-top kernel, and the kernel $k$ used for the smoothed moment indicators. This version of FMB also achieves higher-order correctness since we maintain the asymptotic pivotal nature of the test statistics.

\subsection{From confidence regions to confidence intervals} \label{CICR} 

FMB user can manipulate the higher-order correct CR $\mathcal{C}_{1-\alpha}$ to obtain CR for a subset of parameters, or CI for a single parameter. In this section, we will consider separately the (possibly multivariate) parameter of interest $\beta^{(1)}_0,$ and the (possibly multivariate) nuisance parameter $\beta^{(2)}_0.$ Without loss of generality, we write the partition in the order $\beta = (\beta^{(1)\intercal},\beta^{(2)\intercal})^{\intercal}.$

We define the CR for $\beta^{(1)}_0$ as $\mathcal{C}^{(1)}_{1-\alpha} := \{ \beta^{(1)} : \hat{Q}(\beta^{(1)},\hat{\beta}^{(2)}) \leq q^{\ast}_{1-\alpha}\}.$ This manipulation preserves higher-order correctness when $\hat{Q} ( \beta^{(1)}_0, \hat{\beta}^{(2)} ) \leq \hat{Q} ( \beta_0 )$ $a.s.$, in the sense that it ensures $\mathbb{P}[ \beta^{(1)}_0 \in \mathcal{C}^{(1)}_{1-\alpha}] = \mathbb{P}[ \hat{Q} ( \beta^{(1)}_0, \hat{\beta}^{(2)} ) \leq {q}^{\ast}_{1-\alpha}] \geq \mathbb{P}[ \hat{Q} ( \beta_0 ) \leq {q}^{\ast}_{1-\alpha}] = \mathbb{P}^{\ast} [ Q^{\ast} \leq {q}^{\ast}_{1-\alpha}] + R_T = 1-\alpha + R_T.$ This condition is satisfied, for instance, when $\hat{Q}$ is a monotonic function of the norm $\|\beta-\hat{\beta}\|,$ since $\| (\beta_0^{(1)\intercal}, \hat{\beta}^{(2)\intercal})^{\intercal} - \hat{\beta} \| \leq  \| \beta_0 - \hat{\beta} \|$. It is also satisfied when the two sets of parameters $\beta^{(1)}$ and $\beta^{(2)}$ can be estimated independently from each other. For instance, in a two-step OLS estimation, it is the case in our ACD(1,0) example of Section \ref{Monte_Carlo}, by the Frisch-Waugh-Lovell Theorem, since we can rewrite the ACD(1,0) as an AR(1) process with orthogonal regressors.

If the inequality $\hat{Q} ( \beta^{(1)}_0, \hat{\beta}^{(2)} ) \leq \hat{Q} ( \beta_0 )$ fails to be true almost surely, we cannot guarantee the higher-order correctness of $\mathcal{C}^{(1)}_{1-\alpha}$. Nevertheless, the inequality $\mathbb{P}[ \hat{Q} ( \beta^{(1)}_0, \hat{\beta}^{(2)} ) \leq {q}^{\ast}_{1-\alpha}] \geq \mathbb{P}[ \hat{Q} ( \beta_0 ) \leq {q}^{\ast}_{1-\alpha}]$ remains true for large enough $T,$ as long as $\hat{Q} ( \beta^{(1)}_0, \hat{\beta}^{(2)} ) \overset{\mathcal{D}}\rightarrow \mathcal{X}_{\dim(\beta^{(1)}_0)}$ and $\hat{Q} ( \beta_0 ) \overset{\mathcal{D}}\rightarrow \mathcal{X}_{p}$. This general result implies  that $\mathcal{C}^{(1)}_{1-\alpha}$ contains $\beta^{(1)}_0$ at least with probability $1 - \alpha$ asymptotically, but not always with higher-order refinements.

The drawback of  no guarantee of higher-order refinements is inherent to the existing information on a multivariate parameter. The difficult task of reducing a CR for $\beta$ to a CR for $\beta^{(1)}$ is not directly entangled with the FMB, but more with the nature of dependence between $\beta^{(1)}$ and $\beta^{(2)}$ induced by the moment condition themselves. However, there exists a general way to build CR for $\beta_0^{(1)}$ while preserving higher-order refinements.  Indeed, defining the profile statistic $\bar{Q}(\beta^{(1)}):=\inf_{\beta^{(2)}}\hat{Q}(\beta^{(1)},\beta^{(2)}),$ we get $\bar{Q}(\beta^{(1)}_0)\leq \hat{Q} ( \beta_0 )$ $a.s.$  by construction. Thus, by the same argument, the alternative CR $\bar{\mathcal{C}}^{(1)}_{1-\alpha} := \{ \beta^{(1)} : \bar{Q}(\beta^{(1)}) \leq q^{\ast}_{1-\alpha} \}$ preserves the higher-order refinements. This property comes with a cost, as $\bar{\mathcal{C}}^{(1)}_{1-\alpha}$ is generally more conservative than $\mathcal{C}^{(1)}_{1-\alpha}$ and it might be heavy to compute in high dimension.

\subsection{From confidence intervals to a confidence curve} \label{CICV}

Let us now make connections to the concept of Confidence Distributions (CD) that we use in our empirical application. It aims at answering the following question: can we also use a distribution function, or a “distribution estimator”, to estimate or test for a parameter of interest
in frequentist inference in the style of a Bayesian posterior? (see the review paper by \cite{singh_confidence_2013}). Natural point estimators include the median, the mean, and the maximum of the CD density (\cite{singh_combining_2005}).
That “distribution estimator” is named CD in agreement with the terminology coined by \cite{efron_fisher_1998}, and traces back to the fiducial distribution of \cite{fisher_inverse_1930}, albeit being a purely frequentist concept.
It was introduced by \cite{hjort_confidence_2002} and its asymptotic extension by \cite{singh_combining_2005}
(see also \cite{singh_confidence_2011}, \cite{veronese_fiducial_2015}, and the book-length presentation of \cite{hjort_confidence_2016}).
Example 2.4 of \cite{singh_combining_2005} discusses how a bootstrap distribution can yield a valid asymptotic CD, and Section 2.3.3 of \cite{singh_confidence_2013} how studentization can transmit higher-order accuracy in the i.i.d.\ case. Paralleling these recent developments in fiducial inference theory (see also the review paper of \cite{hannig_generalized_2016}), we can exploit our FMB to produce a fast methodology to build an asymptotically higher-order correct CD as a by-product. Let us define the functions $H_{S}(\beta):=\mathbb{P}[\hat{S}\leq \hat{S}(\beta)]$ and its FMB counterpart $H^{\ast}_{S}(\beta):=\mathbb{P}^{\ast}[S^{\ast} \leq \hat{S}(\beta)]$, for $\beta \in \mathcal{B}$.

\begin{corollary}\label{CDS}
Under Assumptions \ref{a1}---\ref{EE5}, $\mathbb{E} \left[ \left\| g_t \right\|^{\bar q s + \delta} \right] < \infty$, for $\delta > 0$, $s \geq 8$, and $\bar q \geq 3$, $\log T = o(B_T)$, we have the uniform error bound: $\sup_{\beta \in \mathcal{B}}\left\lvert H^{\ast}_{S}(\beta) - H_{S}(\beta)\right\rvert = o_p\left(T^{-1/2}\right) + O_p\left(B_T^{-q}\right) + O_p\left(B_T/T \right),$ and $H^{\ast}_{S}(\beta)$ is an asymptotic CD.
\end{corollary}

We omit the proof since the uniform error bound follows immediately from the proof of Theorem \ref{higher_order}. The second statement comes from the two conditions of Definition 1.1. of \cite{singh_combining_2005} being met, namely $H^{\ast}_{S}(\beta)$ is a cdf, and $H^{\ast}_{S}(\beta_0)$ is uniformly distributed on the unit interval when $T$ goes to infinity. Here, as clarified by \cite{pitman_statistics_1957}, we follow indeed the frequentist view. In $H_{S}(\beta)$ and $H^{\ast}_{S}(\beta)$, randomness is not coming from the (non-random) parameter $\beta$, but from $\hat{S}$ and $S^{\ast}.$

As described in \cite{singh_combining_2005} (see also \cite{fraser_fiducial_1961}, \cite{singh_confidence_2013}), we can also use CD to get $p$-values. For example, the classical bootstrap $p$-value of $H_0 : \beta_0 \leq \beta$ versus $H_1: \beta_0 > \beta$ corresponds to $H^{\ast}_{S}(\beta)$, and the classical equal-tail bootstrap $p$-value of $H_0: \beta_0 = \beta$ versus $H_1: \beta_0 \neq \beta$ corresponds to $2 \min \{H^{\ast}_{S}(\beta), 1-H^{\ast}_{S}(\beta)\}$.

These $p$-values also benefit from higher-order correctness. Collecting them for different values of $\beta$ yields the so-called confidence curve $CV^{\ast}(\beta):=2 \min \{H^{\ast}_{S}(\beta), 1-H^{\ast}_{S}(\beta)\}$, introduced by \cite{birnbaum_confidence_1961} (see \cite{singh_confidence_2013} and \cite{hannig_generalized_2016}  for illustrations). We can view this graphical tool as a piled-up form of two-sided CI of equal tails, at all confidence levels. We provide an example of such a plot in Figure \ref{CC1} for our empirical application in Section \ref{empirical_application}, where we compare CI given by our FMB and first-order Gaussian asymptotics. Finally, \cite{coudin_finite-sample_2020} show how we can design a Hodges-Lehmann-type point estimator (\cite{hodges_estimates_1963}) when a CD is constructed from a hypothesis test.

There exist analogue multivariate CD, for instance Definitions 5.1 and 5.2 in \cite{singh_confidence_2007}. Similarly to the univariate case, we can apply FMB to achieve higher-order accuracy, as long as these multivariate confidence distributions are based on the test statistic $\hat{Q}(\beta_0)$.

\section{Monte Carlo experiments}\label{Monte_Carlo}

To illustrate the applicability of FMB, we consider a simulation exercise on constructing CR for the parameters of an Autoregressive Conditional Duration (ACD) model (\cite{engle_autoregressive_1998}). It is a model typically applied for the analysis of high-frequency data in finance, and more generally to model positive variables (e.g. volatility or volume) via a multiplicative error model; see e.g.\ \cite{hautsch_econometrics_2012} for a recent book-length presentation.

The duration is defined as the time lag  between two consecutive events occurrence, namely $x_\ell := t_\ell-t_{\ell-1}$. Clearly, $x_\ell >0 $, for any $\ell\in \mathbb{T}$. We model $\mathbb{E} \left( x _ { \ell } | x _ { \ell - 1 } , \ldots , x _ { 1 } \right) = m _ { \ell } \left( x _ { \ell - 1 } , \ldots , x _ { 1 } ; \beta \right) := m _ { \ell }$, assuming the 
model $x_\ell = m_\ell \varepsilon_\ell$, with $\epsilon_\ell \overset{i.i.d.} \sim \mathcal{E}\left( 1 \right)$ for any $\ell$, with $\mathcal{E}\left( 1 \right)$ being an exponential random variable with mean one. Specifically, for the ACD(1, 1) specification, we have 
\begin{equation} \label{Eq: ACD}
x_\ell = \epsilon_\ell m_{\ell}, \text{\ \ with \ \ } m _ { \ell } = \omega + \beta_1 x _ { \ell - 1 } + \beta_2 m _ { \ell - 1 }  ,\qquad \ell\in\mathbb{Z},
\end{equation}  
for $\omega > 0$  $\beta_1 , \beta_2 \in \mathbb{R}^{+}$ and  $\beta_1 + \beta_2 < 1$.
When we take $\beta_2 = 0$, the ACD(1,0) model is in agreement with Assumptions \ref{a1}--\ref{EE5} (in Section \ref{Theory}). Thus, we start our numerical experiment with this specification, conducting inference on $\beta:=(\omega,\beta_1)^{\intercal}$. We apply the optimal estimating functions of \cite{li_semiparametric_2000} and a moment condition which does not assume any specific functional form for the underlying innovation density, but relies on the 
unconditional expectation of $x_\ell$. Therefore, given a random sample of durations 
$(x_1,...,x_\ell,...,x_T)$, the  vector of moment 
conditions for the $\ell$-th observation is $g_\ell ( \beta ) := \left( g_{1,\ell}(  \beta), g_{2,\ell}
(\beta), g_{3,\ell}( \beta)\right)^{\intercal},$
with $g_{1,\ell}\left( \beta\right) := (\left( x_\ell - m_\ell \right) / m _ { \ell } ^ { 2 }  )(  \partial m _ { \ell } / \partial \omega  )$, $g_{2,\ell}\left( \beta \right) := (\left( x_\ell - m_\ell \right) / m _ { \ell } ^ { 2 }  )(  \partial m _ { \ell } / \partial \beta_1  )$, and $g_{3,\ell}\left( \beta\right) := x_\ell  - \left( 1-\beta_1 \right)^{-1}$ (\cite{li_semiparametric_2000}). Thus, we are in the over-identified case with $r=3$ for $p=2$.

In a second step, to get numerical insights on the applicability of our FMB, we extend our Monte Carlo experiment to a general ACD(1,1) model. The latter is non-markovian in the observations and does not fit our setting stricto sensu, since we assume the vector of observations in \eqref{moment} to be finite to ease notations and proofs. However, taking a large number $v$ of lagged durations in an ACD($v$,0) model is close to an ACD(1,1) model, and it meets our current theoretical framework.

We conduct inference on $\beta:=(\omega,\beta_1,\beta_2)^{\intercal}$ and our moment 
conditions for the $\ell$-th observation is $g_\ell ( \beta ) := \left( g_{1,\ell}(  \beta), g_{2,\ell}
(\beta), g_{3,\ell}( \beta), g_{4,\ell}( \beta) \right)^{\intercal},$ with $g_{1,\ell}\left( \beta\right) := (\left( x_\ell - m_\ell \right) / m _ { \ell } ^ { 2 }  )(  \partial m _ { \ell } / \partial \omega  )$, $g_{2,\ell}\left( \beta \right) := (\left( x_\ell - m_\ell \right) / m _ { \ell } ^ { 2 }  )(  \partial m _ { \ell } / \partial \beta_1  )$, 
$g_{3,\ell}\left( \beta \right) := (\left( x_\ell - m_\ell \right) / m _ { \ell } ^ { 2 }  ) (  \partial m _ { \ell } / \partial \beta_2  ),$ 
and $g_{4,\ell}\left( \beta\right) := x_\ell  - \left( 1-\beta_1-\beta_2 \right)^{-1}$. We stay in the over-identified case with $r=4$ for $p=3$.

In the following, we compute by Monte Carlo simulations the coverage of FMB CR. We label the results $\hat{Q}_{FMB}$ for the FMB using $\hat{Q},$ $\tilde{Q}_{FMB}$ for the FMB using the cubic approximation of Remark \ref{nomono2} and $\tilde{Q}_{FMB,2}$ for the FMB using the quadratic  approximation of Remark \ref{nomono2}. To validate numerically our theoretical results, we compare the coverage of these FMB CR to some standard first-order correct alternatives. The first one, labeled as $\hat{Q}_{\mathcal{X}^2_r}$, defines CR as contours of the same statistic than FMB, but making use of the (first-order correct) $\mathcal{X}^2_{r}$ asymptotic distribution to compute the rejection probabilities. The second one is the standard elliptical contour of an asymptotically $\mathcal{X}^2_{p}$ distributed Wald statistic (labeled as $W_{\mathcal{X}^2_{p}}$), whose covariance matrix is a HAC estimator with bandwidth $B_T$.

To compare with a state-of-the-art competitor of FMB in terms of higher-order correctness, we also show the coverage of CR yielded by MBB. We adapt the MBB of \cite{inoue_bootstrapping_2006} to a Wald statistic, which is, in turn, an adaptation of \cite{gotze_second-order_1996}. 
As the statistic defining the CR has to be asymptotically pivotal to obtain higher-order correctness, we choose to apply MBB to the Wald statistic, which we label as $W_{MBB}.$ We show a detailed CPU time comparison between FMB and MBB in the Supplementary Material (Table \ref{Tab: CPU}). In general, FMB appears to be at least $10^3$ times faster than MBB as expected.

In Table \ref{Tab: 10_025COVER_3}, we display the empirical coverages for the ACD(1,0) model, for $B_T=3$ and $B_T=5.$ In Table \ref{Tab: 025COVER_3}, we show the same outputs for the ACD(1,1) specification.

\begin{table}[htbp]
\begin{center}
\caption{Empirical coverage of CR, a comparison between first and higher-order correct methods.}
\begin{tabular}{l l l l l l l l}
\hline
\hline

	\\

  &&\multicolumn{3}{c}{ $B_T=3$ } & \multicolumn{3}{c}{$B_T=5$} \\
	
  \multicolumn{2}{l}{Coverages:}  & 0.90 & 0.95 & 0.99   & 0.90 & 0.95 & 0.99 \\
	\\
  \multirow{6}{*}{\rotatebox[origin=c]{90}{$T=250$}} & $\hat{Q}_{FMB}$ & 0.92 & 0.94 & 0.98   & 0.89 & 0.93 & 0.96 \\ 

  &$\tilde{Q}_{FMB}$ & 0.92 & 0.95 & 0.98 & 0.90 & 0.94 & 0.97 \\
	&$\tilde{Q}_{FMB,2}$  & 0.92 & 0.95 & 0.98 & 0.90 & 0.93 & 0.97  \\
	&$\hat{Q}_{\mathcal{X}^2_r}$ & 0.91 & 0.93 & 0.97 & 0.88 & 0.90 & 0.96 \\ 
  &$W_{MBB}$  & 0.93 & 0.97 & 1.00 & 0.93 & 0.97 & 1.00 \\ 
  &$W_{\mathcal{X}^2_{p}}$  & 0.84 & 0.90 & 0.96 & 0.80 & 0.86 & 0.92 \\ 
	\\
	
   \multirow{6}{*}{\rotatebox[origin=c]{90}{$T=500$}} & $\hat{Q}_{FMB}$ & 0.93 & 0.96 & 0.98 & 0.92 & 0.95 & 0.98  \\ 

  & $\tilde{Q}_{FMB}$  & 0.93 & 0.95 & 0.98 & 0.92 & 0.95 & 0.98 \\
	& $\tilde{Q}_{FMB,2}$  & 0.92 & 0.96 & 0.99 & 0.91 & 0.95 & 0.99 \\
	& $\hat{Q}_{\mathcal{X}^2_r}$ & 0.92 & 0.95 & 0.98 & 0.91 & 0.94 & 0.97 \\ 
  & $W_{MBB}$ & 0.95 & 0.98 & 1.00  & 0.94 & 0.98 & 1.00 \\ 
  & $W_{\mathcal{X}^2_{p}}$ & 0.86 & 0.91 & 0.97  & 0.84 & 0.89 & 0.96 \\ 
	\\
  \hline
	\hline
\end{tabular}
\label{Tab: 10_025COVER_3}
\caption*{

The true values of the unknown parameters of the ACD(1,0) are $\omega = 1.5$ and $\beta_1 = 0.25$.}
\end{center}
\end{table}

\begin{table}[htbp]
\begin{center}
\caption{Empirical coverage of CR, a comparison between first and higher-order correct methods.}
\begin{tabular}{l l l l l l l l}
\hline
\hline
\\

&&\multicolumn{3}{c}{$B_T = 3$}&\multicolumn{3}{c}{$B_T = 5$} \\

\multicolumn{2}{l}{Coverages:} & 0.90 & 0.95 & 0.99 & 0.90 & 0.95 & 0.99 \\
\\

 \multirow{6}{*}{\rotatebox[origin=c]{90}{$T=250$}} & $\hat{Q}_{FMB}$  & 0.88 & 0.93 & 0.98  & 0.87 & 0.90 & 0.96\\ 
    & $\tilde{Q}_{FMB}$  & 0.89 & 0.92 & 0.95 & 0.87 & 0.90 & 0.94 \\
		& $\tilde{Q}_{FMB,2}$  & 0.82 & 0.87 & 0.94 & 0.80 & 0.85 & 0.93 \\
		& $\hat{Q}_{\mathcal{X}^2_r}$  & 0.86 & 0.91 & 0.96 & 0.85 & 0.89 & 0.95 \\
    & $W_{MBB}$  & 0.97 & 0.99 & 1.00 & 0.97 & 0.99 & 1.00 \\ 
    & $W_{\mathcal{X}^2_{p}}$  & 0.73 & 0.79 & 0.88 & 0.71 & 0.77 & 0.86 \\ 
		\\
   \multirow{6}{*}{\rotatebox[origin=c]{90}{$T=500$}} & $\hat{Q}_{FMB}$ & 0.91 & 0.95 & 0.99 & 0.90 & 0.94 & 0.98\\ 
    & $\tilde{Q}_{FMB}$ & 0.91 & 0.94 & 0.97 & 0.91 & 0.94 & 0.97 \\
		 & $\tilde{Q}_{FMB,2}$ & 0.85 & 0.90 & 0.96 & 0.84 & 0.89 & 0.95 \\
		& $\hat{Q}_{\mathcal{X}^2_r}$  & 0.90 & 0.93 & 0.97 & 0.89 & 0.93 & 0.97 \\
    & $W_{MBB}$  & 0.98 & 0.99 & 1.00  & 0.97 & 0.99 & 1.00 \\ 
    & $W_{\mathcal{X}^2_{p}}$ & 0.78 & 0.83 & 0.92 & 0.77 & 0.83 & 0.92 \\
	 \\
   \hline
	 \hline

\end{tabular}
\label{Tab: 025COVER_3}
\caption*{

The true values of the unknown parameters of the ACD(1,1) are $\omega=1.5,$ $\beta_1 = 0.25$ and $\beta_2 = 0.25$.}
\end{center}
\end{table}

\newpage

In line with our theoretical results, we observe that FMB performs well compared to its first-order correct competitors: the coverages are typically closer to their nominal level. It generally remains true for the $\tilde{Q}_{FMB}$ version of FMB CR, whose coverage is often very close to the one of the original FMB CR based on $\hat{Q}_{FMB}.$ The coverage of the quadratic $\tilde{Q}_{FMB,2}$ version of FMB CR is generally further away from the original FMB.

The Wald statistic seems to yield very erratic CR (which is additionally confirmed by unreported plots). The MBB version of the Wald statistic improves slightly on the asymptotic $\mathcal{X}^2_p$ distribution, without being convincing though.

Finally, the FMB presented here does not take advantage of all the potential fine-tuning, and this should leave room for practical improvement. First, we might improve FMB if the long-run variance $\Omega(\beta_0)$ is estimated with a less biased version of variance estimator, for instance carrying out a prewhitening step (\cite{andrews_improved_1992}), or using a flat-top kernel (\cite{politis_higher-order_2011}) as discussed in Section \ref{sectionHAC}. It should yield a smaller bootstrap error, as shown in Theorem \ref{higher_order}. Second, we stress that the moment indicators do not have necessarily the same dependence structure. Thus, smoothing the multivariate moment indicators with different bandwidths might further improve the coverage of FMB.

\section{Real data application}\label{empirical_application}

In this section, we illustrate how FMB performs on real data. We look at daily volumes of stock transaction (in millions), modeled with the same exponential ACD as in Section \ref{Monte_Carlo} (see \eqref{Eq: ACD}). We focus on data available online (\textit{Yahoo! Finance}), for five stocks 
in three different sectors, namely bank, technology, and food. We compute the CR (for parameters $\omega$, $\beta_1$ and $\beta_2$), before  the subprime crisis (2005), during the crisis (2008), and the current period (2018). The sample size of each period corresponds to the number of trading days, namely $T=252$ up to negligible variations from year to year. Before diving into a deeper analysis, we briefly describe the data at hand in Table \ref{Tab: SUM} below.

\begin{table}[htbp]
\begin{center}
\caption{Summary statistics of volumes of transaction.}
\begin{tabular}{|lllllllll|}
\hline
    \textit{Year 2005} & Min & Max & Med & Mean & IQR & SD & SKN & KURT \\
\hline
 \rowcolor{mygray} BA & 4.52 & 42.05 & 10.67 & 11.24 & 4.98 & 4.47 & 2.36 & 11.18 \\

JPM & 3.77 & 25.47 & 9.93 & 10.6 & 4.11 & 3.25 & 0.94 & 1.31 \\

 \rowcolor{mygray} MSF & 27.21 & 187.38 & 63.53 & 66.61 & 20.94 & 20.23 & 1.78 & 6.40 \\

KO & 3.96 & 39.75 & 11.01 & 11.86 &  3.83 & 4.15 & 2.64 & 12.16\\

 \rowcolor{mygray} UL & 0.16 & 3.36 & 0.49 & 0.59 & 0.31 & 0.39 & 2.91 & 12.78\\

\hline \hline
\textit{Year 2008} & Min & Max & Med & Mean & IQR & SD & SKN & KURT \\ 
\hline 

 \rowcolor{mygray} BA & 22.59 & 322.73 & 63.16  & 78.63 & 59.52 & 49.01 &  1.66 & 3.28\\

 JPM & 12.34 & 194.07   & 41.96 & 48.99 & 29.64 & 25.90 & 1.93 & 5.72\\

 \rowcolor{mygray} MSF & 16.88 & 291.14  & 78.50  & 84.17 & 38.81 & 35.51 & 1.55 & 4.73 \\

 KO & 5.32 & 79.21 & 23.35 & 25.26 & 12.77 & 10.63 & 1.62 & 4.20\\

 \rowcolor{mygray} UL & 0.26 & 5.19 &  0.79  & 1.08 & 0.73 & 0.83 & 2.09 & 4.85\\

\hline \hline
  \textit{Year 2018}  & Min & Max & Med & Mean & IQR & SD & SKN & KURT \\
\hline
 \rowcolor{mygray} BA & 22.97 & 165.88 & 62.21 & 67.91 & 29.16 & 24.15 & 1.29 & 2.02\\

 JPM & 6.49 & 41.31 & 13.90 & 15.17 & 6.43 & 5.53 & 1.60 & 3.63\\

 \rowcolor{mygray} MSF & 13.66 & 111.24 & 27.61 & 31.59 & 14.23 & 13.40 & 1.74 & 4.82\\

 KO &  4.79  & 32.48 & 11.91  & 12.52 & 4.32 & 4.13 & 1.32 & 2.67\\

 \rowcolor{mygray} UL & 0.33 & 4.88 & 0.90  & 1.05  & 0.54 & 0.64 & 2.96 & 11.54\\

\hline
\end{tabular}
\caption*{ 

We consider the companies Bank of America (BA), JP Morgan (JPM), Microsoft (MSF), Coca-Cola (KO) and Unilever (UL). In the summary, Med stands for the median, IQR for the inter-quantile range, SD for the standard deviation, SKN for skewness, and KURT for excess of kurtosis.}
\label{Tab: SUM}
\end{center}
\end{table}
Table \ref{Tab: SUM} illustrates the larger variability of the volumes of transaction during 2008, as measured by the standard deviation (SD) and the interquartile range (IQR). The high skewness (SKN) and excess of kurtosis (KURT) typically indicate that a higher-order correct inferential procedure might be required in finite samples.

To investigate further the impact that asymmetry and fat tails may have on the conducted inference, we compute the FMB and the asymptotic normal (Asy) CR of nominal coverage $(1-\alpha) = 95\%$, for the ACD(1,1) parameters $\omega,$ $\beta_1$ and $\beta_2$ at each period. As FMB yields higher-order accurate CR by inverting probabilities of the test statistic $\hat{Q}(\beta_0)=\hat{Q}(\omega_0,\beta_{1,0},\beta_{2,0})$, we represent these trivariate CR by slicing them at the estimates $\hat{\omega},$ $\hat{\beta}_1$ and $\hat{\beta}_2$ (Table \ref{Tab: Big1}). Namely, we cut the CR by fixing the parameters that are not of interest to their estimated values.
To keep Table \ref{Tab: Big1} concise, we do not report the estimate $\hat{\omega}$ and the intervals for the parameter $\omega$. We can deduce the former from Table \ref{Tab: SUM}.\footnote{Using Section \ref{Monte_Carlo} and volumes of transaction $\{x_t\}_{t=1}^{T}$ instead of durations, we have $\hat{\omega} \approx (1-\hat{\beta_1}-\hat{\beta_2})T^{-1}\sum_{\ell=1}^{T}x_{\ell},$ from the moment condition based on $g_{3,t}.$ We report the sample mean $T^{-1}\sum_{\ell=1}^{T}x_{\ell}$ in Table \ref{Tab: SUM}.}

\begin{table}[htbp]
\begin{center}
\small\addtolength{\tabcolsep}{-5pt}
\caption{Analysis of volumes of transaction.}
\begin{tabular}{|c|c|cc|cc|cc|}
\hline
\multicolumn{2}{|c|}{\multirow{2}{*}{Asset}} & \multicolumn{2}{|c|}{2005} & \multicolumn{2}{|c|}{2008} & \multicolumn{2}{|c|}{2018} \\ \cline{3-8}
\multicolumn{2}{|c|}{}                        & \multicolumn{1}{|c|}{$\hat{\beta_1}$}   & \multicolumn{1}{|c|}{$\hat{\beta_2}$}  & \multicolumn{1}{|c|}{$\hat{\beta_1}$}  & \multicolumn{1}{|c|}{$\hat{\beta_2}$}  & \multicolumn{1}{|c|}{$\hat{\beta_1}$}  & \multicolumn{1}{|c|}{$\hat{\beta_2}$} \\ \hline

\multirow{3}{*}{JPM}         & Est            &   0.402   &   0.132   &    0.691    &   0.122     &   0.459   &   0.345   \\
                             & Asy            &    [0.384,0.420]     &    [0.114,0.150]      &    [0.673,0.709]    &   [0.104,0.140]   &   [0.448,0.469]   &  [0.335,0.355]   \\
                             & FMB        &    [0.357,0.442]    &   [0.085,0.178]     &    [0.621,0.726]    &   [0.060,0.164]     & [0.429,0.483]    &    [0.318,0.372]  \\ \hline

\multirow{3}{*}{BA}          & Est         &     0.366    &  0.534    &    0.624    &   0.323   &  0.538  &   0.141  \\
                             & Asy           &     [0.361,0.371]    &    [0.529,0.540]    &   [0.618,0.629]   &   [0.318,0.328]   &  [0.521,0.554]   &   [0.125,0.157]  \\
                             & FMB           &     [0.345,0.381]     &   [0.512,0.549]     &    [0.594,0.641]    &   [0.295,0.344]    & [0.491,0.584]      &    [0.102,0.192]   \\ \hline

\multirow{3}{*}{MSF}         & Est          &   0.271   &  0.340     &   0.595  &   0.272      &  0.570   &  0.268  \\
                             & Asy         &   [0.261,0.281]   & [0.330,0.350]   &    [0.586,0.604]   &   [0.262,0.281]     &  [0.559,0.582]   & [0.257,0.280]      \\
                             & FMB          &  [0.242,0.297]   &  [0.307,0.373] &  [0.562,0.623]  &  [0.243,0.301]   &   [0.536,0.596]   & [0.238,0.294]    \\	\hline

\multirow{3}{*}{KO}         & Est         &    0.202    &   0.377      &   0.488    &    0.371     &   0.392    &   0.382   \\
                             & Asy           &     [0.190,0.213]       &  [0.365,0.389]    &   [0.479,0.496]    &   [0.363,0.380]     &   [0.384,0.399]   &    [0.375,0.390]    \\
                             & FMB         &   [0.166,0.233]  &   [0.344,0.426]     &  [0.458,0.513]    &  [0.343,0.400]    &    [0.363,0.420]    &    [0.358,0.413]   \\ \hline

\multirow{3}{*}{UL}         & Est          &   0.331   &  0.529    &   0.571  &   0.290      &  0.435  &   0.483 \\
                             & Asy          &    [0.320,0.343]     &  [0.518,0.541]    &    [0.555,0.586]   &   [0.275,0.305]     &   [0.428,0.441]  &   [0.477,0.490]  \\
                             & FMB         &  [0.297,0.365]  &   [0.509,0.571]    &  [0.513,0.599]  &  [0.236,0.321]   &   [0.413,0.456]   & [0.464,0.503]     \\	\hline

\end{tabular}
\label{Tab: Big1}
\caption*{ 

We consider the companies Bank of America (BA), JP Morgan (JPM), Microsoft (MSF), Coca-Cola (KO) and Unilever (UL). We use the Exponential Tilting estimator (Est), a particular case of GEL, and the benchmark intervals are the first-order asymptotic Gaussian (Asy). All the intervals are equal-tailed and have conditional nominal coverage of $95\%$ when we fix the other parameters at their estimated values.}
\end{center}
\end{table}
A few comments are in order. First of all, the different sectors exhibit very diverse reactions to the events happening in 2008. For instance, the food sector seems to be the most stable, while financial sector undergoes a huge variability, as we could expect. We can observe this  either comparing non-critical periods to the crisis, or comparing the estimates and their CI before and after the crisis. For instance, the estimates for Unilever are almost the same before and after the crisis, as if the company has recovered the same volume behaviour. Coca-Cola looks equally stable with respect to the parameter $\beta_2$, which is almost unchanged after the crisis. Second, the estimate $\hat{\beta_1},$ respectively $\hat{\beta_2},$ seems to be larger, respectively smaller, during the crisis period.  It is expected since $\beta_1$ reflects the sudden trading reactions due to changes in the expectations by the market participants during the crisis period. Thus, this feature of 2008 corresponds to an increase of the impact of news (shocks) on the volumes of transaction (via the parameter $\beta_1$), relative to persistence (via the parameter $\beta_2$). Finally, we see that the FMB CI are longer than the first-order correct Gaussian CI. It is in line with our Monte Carlo experiments, as available in Section \ref{Monte_Carlo}. Indeed, as the CR are defined by level sets of $\hat{Q}$, a longer CI corresponds to an adaptation of FMB to a skewed or fat-tailed distribution. Since the CI obtained by Gaussian approximation are typically shorter, we conclude that the routinely applied first-order asymptotic theory tends to underestimate the rejection probability, whereas FMB stays conservative.
Our experience underpinned by several Monte Carlo simulations makes us expect that the distribution of $\hat{Q}(\beta_0)$ is more skewed or fat-tailed than the chi-squared; see the comparison between $\hat{Q}_{\mathcal{X}^2_r}$ and FMB in Table \ref{Tab: 025COVER_3}.

\begin{figure}[htbp]
\begin{center}
\caption{Confidence Curves of the FMB and Gaussian approximation.}\label{CC1}
\includegraphics[width = 15cm, height = 11cm]{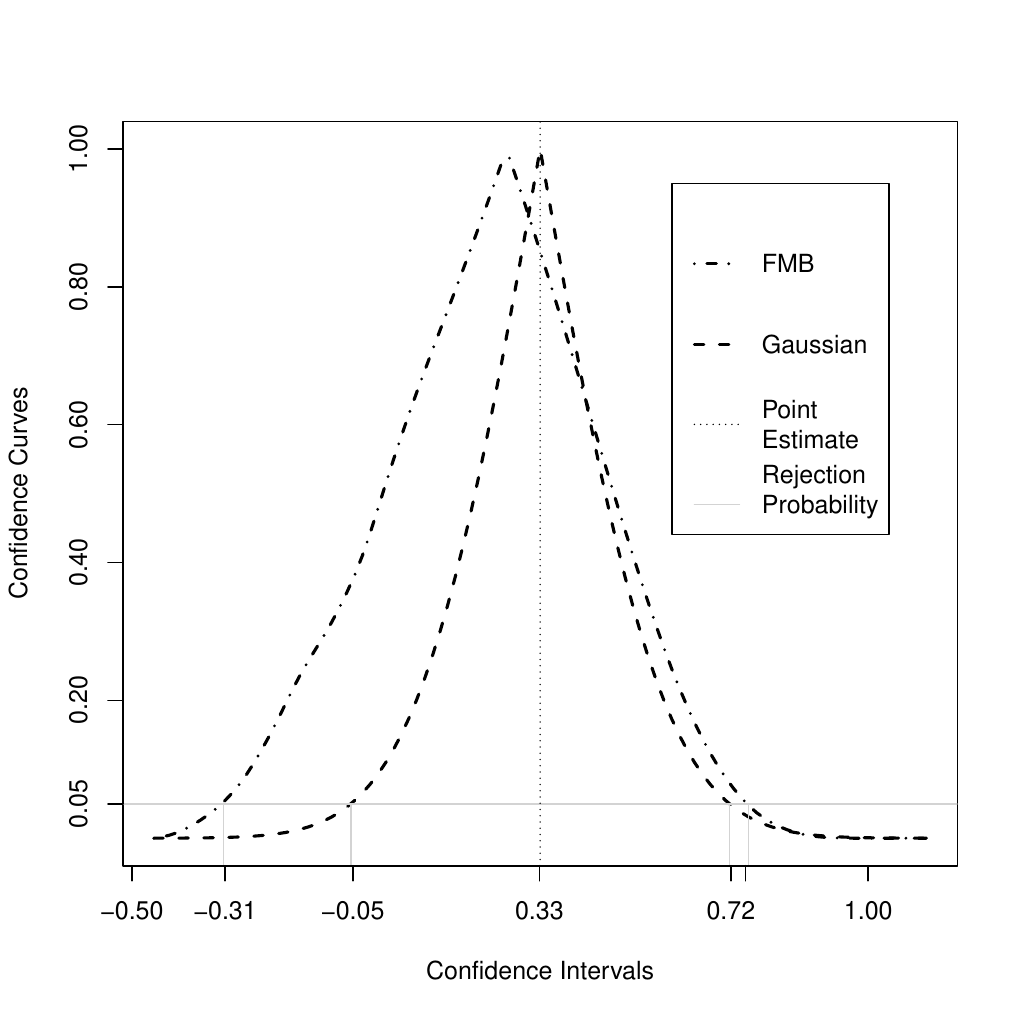}
\caption*{Confidence curve for the parameter $\beta_1$ of Unilever in 2005, with point estimate $\hat{\beta}_1=0.33$. The sample size is $T=252$, and the nominal coverage of the confidence intervals is $(1-\alpha)=95\%$. The flat solid line is the rejection probability level $5\%$, and its crossing with the confidence curves gives the FMB $95\%$ equal-tailed CI $[-0.31, 0.75]$ and Gaussian $95\%$ equal-tailed CI $[-0.05, 0.72]$.}
\end{center}
\end{figure}

Following our discussion on CD (Subsection \ref{CICV}), we illustrate here the link between our FMB CR and our previous definition of asymptotic confidence distribution $H^{\ast}_S(\beta)$, via the confidence curve $CV^{\ast}(\beta)$. Among the alternative ways to represent the former CR, marginalization allows us to build unconditional CI. Stacking the CR at different coverages $1-\alpha$ leads to a center-outward confidence curve for the multidimensional parameter $CV(\omega,\beta_1,\beta_2)$. For each CI, we integrate out the two parameters that are not of interest in $CV(\omega,\beta_1,\beta_2)$. This yields a different confidence curve $CV^{\ast}(\beta)$ for each $\beta \in \{\omega,\beta_1,\beta_2\},$ whose level sets give the equal-tailed CI. As an illustration of graphical use of these confidence curves (defined in Section \ref{Theory}), Figure \ref{CC1} reports a comparison between the FMB and Gaussian CI based on the FMB and Gaussian confidence curves. Again we observe that FMB is much more conservative.

\newpage

\section*{Acknowledgement}
We would like to thank the editor, the co-editor, and the referee for constructive criticism and numerous
suggestions which have led to substantial improvements over the previous versions. We thank A.-P.\ Fortin, E.\ Paparoditis, D.\ Politis, and R.\ J.\ Smith for helpful comments and discussion, as well as participants in the North American Summer Meeting of the Econometric Society (Seattle, 2019),  Geneva Finance Research Institute seminar, Workshop on Statistical Learning (Geneva, 2019-20), online RCEA Time Series Workshop 2021, and online IAAE Conference 2021.


\section*{}
\bibliography{FMB_JOE} 

\newpage

\section*{Appendix: Proofs}\label{Proofs}

In this appendix, we prove the asymptotic refinements of FMB. By construction, the higher-order correctness of FMB CI and CR (for $\beta_0$) entirely hinges on FMB applied to the test statistics $\hat{S}(\beta_0)$ (see \eqref{Eq S_psi}) and $\hat{Q}(\beta_0)$ (see \eqref{Rao_type}). Therefore, it follows from Theorem \ref{higher_order} as in Section \ref{Theory}.

The outline of the proof goes as follows. In Appendix A.1, we derive an Edgeworth expansion for $\hat{S}(\beta_0)$ and $\hat{Q}(\beta_0)$. In Appendix A.2, we derive a similar Edgeworth expansion for the bootstrap counterparts $S^{\ast}$ and $Q^{\ast}.$ In Appendix A.3, we show that the difference between the two Edgeworth expansions is of order $o_p\left(T^{-1/2}\right)   + O_p\left(B_T^{-q}\right) + O_p\left(B_T/T \right)$. The first term in this difference is smaller than the usual $O_p(T^{-1/2})$ of the standard first-order asymptotics (central limit theorem). The improvement is essentially due to FMB being able to approximate accurately the third moment of the statistics, whilst CLT approximates only the first two moments. It yields higher-order correctness.

The second and third term of the difference has the same order than the bias and variance of the variance estimator, which scales $\hat{S}(\beta_0)$ and $\hat{Q}(\beta_0)$.
We get the order $O_p( B_T^{-q} )  + O_p\left(B_T/T \right)$, where $q$ is the Parzen exponent of the induced kernel $k^{\ast}$ (see Section \ref{sectionHAC}). With $B_T = o(T^{1/2})$, $q$ has to be larger than one for the overall error of FMB to be $o_p(T^{-1/2})$. We have to consider this aspect in the choice of the kernel $k$ and of the bandwidth $B_T$ entering the construction of the estimator $\hat{\Omega}$ (see Section \ref{Theory} for further details).

For the sake of exposition, technical lemmas and lengthy derivations are available in the online Supplementary Material.

\subsection*{A.1. Edgeworth expansion of the original sample statistics}\label{Original Sample Edgeworth}

In this section, we derive the Edgeworth expansions of $\hat{S}(\beta_0)$ and $\hat{Q}(\beta_0)$.  We state the following:
\begin{theorem}\label{Ed_Bound0}
Under Assumptions \ref{a1}---\ref{Markov_type}, with $s \geq 8$ and $\log T = o(B_T)$, we get the Edgeworth expansions:
\begin{align}
\Upsilon^{\dag}_{S,T}( x ) &= \Phi ( x ) + T ^ { - 1 / 2 } p _ { 1 } ( x, \mathcal{K}_S^1) \phi ( x ) + (B_T / T ) p _ { 2 } ( x,\mathcal{K}_S^2) \phi ( x ),\label{EE1}
\\
\Upsilon^{\dag}_{Q,T} ( x ) &= F _ { \mathcal{X} _ { r } ^ { 2 } } ( x ) + (B_T / T ) p _ { Q } ( x,\mathcal{K}_Q) f _ { \mathcal{X} _ { r } ^ { 2 } } ( x ).\label{EE2}
\end{align}
\noindent with uniform error bound:
\begin{align}
\sup_{x \in \mathbb{R}}\left\lvert \mathbb{P}\left[\hat{S}\leq x\right] - \Upsilon^{\dag}_{S,T} ( x ) \right\rvert =& o\left(B_T / T\right) + O\left(B_T^{-q}\right) + O\left(B_T/T \right), \label{UEB1}
\\
\sup_{x \in \mathbb{R}_{+}}\left\lvert \mathbb{P}\left[ \hat{Q} \leq x\right] - \Upsilon^{\dag}_{Q,T} ( x ) \right\rvert =& o\left(B_T / T\right) +  O\left(B_T^{-q}\right) + O\left(B_T/T \right). \label{UEB2}
\end{align}
In the expansions, $p_1$ is an even polynomial in $x$, and $p_2$ and $p_{Q}$ are odd polynomials in $x$. These polynomials depend on vectors $\mathcal{K}_{S}^1,$  $\mathcal{K}_{S}^2,$ and $\mathcal{K}_Q,$ respectively containing the first three cumulants of $\hat{S}$ and $\hat{Q}.$ The uppercase $\Phi$ and $F_{\mathcal{X}^2_r}$ are respectively the c.d.f.\ of a $\mathcal{N}(0,1)$ and a $\mathcal{X}^2_r$ random variable, the lowercase $\phi$ and $f_{\mathcal{X}^2_r}$ stand for their densities.
\end{theorem}
To prove the statements, we remark that $\hat{S}(\beta_0)$ bears some similarity to the studentized smooth function of means of \cite{gotze_second-order_1996}. Therefore, our proof mainly relies on the same strategy as in G\"otze and K\"unsch paper. However, we have to discuss two important distinctions.

First, \cite{gotze_second-order_1996} derive an Edgeworth expansion considering a class of studentizing factor of the form $\hat{\varsigma}^2 = \kappa_1^2\sum_{s = 1-T}^{T-1} k^\ast \left( s / B_{T} \right) \hat{\Gamma}_s( \hat{\beta} ),$
 with $\hat{\Gamma}_s( \hat{\beta} ) := T^{-1}\sum_{t = 1}^{T-s} g_t\left( \hat{\beta} \right) g^{\intercal}_{t+s}\left( \hat{\beta} \right),$ as introduced by \cite{andrews_heteroscedasticity_1991}. Our definition of the studentizing factor is different, but asymptotically equivalent, since $$\hat{\sigma}^2 = \kappa_1^2 \sum_{s = 1-T}^{T-1} k^{\ast}_T \left( s/B_T \right) \hat{\Gamma}_s(\hat{\beta}),$$ where $k^{\ast}_T \left( s/B_T \right) := (\kappa_2 B_T)^{-1} \sum_{t = \max\left[ 1-T,1-T+s\right]}^{\min\left[ T-1,T-1+s\right]} k\left( t-s/B_T \right) k\left( t/B_T \right)$ is a consistent approximation of $k^{\ast}\left( s/B_T \right)$ by Riemann sum (\cite{smith_automatic_2005}). In particular, the bias and variance of both studentizing factor have the same order, respectively $O(B_T^{-q})$ and $O(B_T/T)$, where $q$ is the Parzen exponent (see Section \ref{Theory}). As a consequence, both studentizing factors act equivalently on the error bound of the Edgeworth expansion for $\hat{S}(\beta_0)$ (see \eqref{ExpandS}).

Second, from the convolution step of Equation \eqref{convol}, we are interested in $T^{1/2}\bar{g}_T = T^{-1/2}\sum_{t=1}^{T}g_{T,t}$ instead of $T^{1/2}\bar{g}=T^{-1/2}\sum_{t=1}^{T}g_{t}$, as in \cite{gotze_second-order_1996}. Thus, we have to derive a valid Edgeworth expansion under this modification. To this end, we rewrite 
\begin{eqnarray*}
T^{1/2}\bar{g}_T &=& T^{-1/2}\sum_{t=1}^{T} B_{T}^{-1/2}\sum_{s = t-T}^{t-1} k\left(s / B_{T}\right) g_{t-s} \qquad = \qquad  T^{-1/2}\sum_{t=1}^{T}g_t B_T^{-1/2}\sum_{s=1-t}^{T-t}k(s/B_T).
\end{eqnarray*}
In this representation, the kernel smoothing induces a tapering window 
$w(t) := B_T^{-1/2}\sum_{s=1-t}^{T-t}k(s/B_T)$ on the summand time series in such a way that $T^{1/2}\bar{g}_T = T^{-1/2}\sum_{t=1}^{T} w(t)g_t$. Hence, we need to check that the regularity conditions given by \cite{gotze_asymptotic_1994} hold for the tapered moment indicators, when they are assumed to be true for the original ones (Assumptions \ref{a1}---\ref{Markov_type}).

\textbf{Proof of Theorem \ref{Ed_Bound0}}.

To check Assumptions \ref{a1}---\ref{Markov_type} for $\left\{w(t)g_t\right\}$, we label as ``RC $i$'' the regularity conditions defined in Assumptions $i$; for instance ``RC \ref{a1}'' represents the regularity condition of Assumption 1. 

When we assume that  RC \ref{a1} and RC \ref{a2} hold for the process $\left\{g_t\right\}$, we can easily verify that they are true also for $\left\{w(t)g_t\right\}$, given that $w(t)<\infty,$ $\forall t$. To check RC \ref{RV_approx}, consider that 
$$
\mathbb{E}\left\| w(t)g_t - w(t)g_{t,m}^{\ddagger} \right\| \leq \lvert \max_{t=1,...,T} w(t) \rvert \mathbb{E} \left\| g_t - g_{t,m}^{\ddagger} \right\| \leq \lvert \max_{t=1,...,T} w(t) \rvert \delta^{-1} \exp\left(-\delta m\right)
$$ 
by Assumption \ref{RV_approx} on $\left\{g_t\right\}$. Thus, the exponential rate of decay of the approximation error is not affected by tapering, and we can always take the process $\{w(t)g_{t,m}^{\ddagger}\}$ to approximate $\{w(t)g_t\}$. Without loss of generality, let us take $\mathcal{D}_t:=\sigma\langle g_{t,0}^{\ddagger}\rangle$, and note that both $w(t)g_{t,0}^{\ddagger}$ and $g_{t,0}^{\ddagger}$ are $\mathcal{D}_t$-measurable. Then, RC \ref{Rosenblatt_mix} and RC \ref{Markov_type} follow immediately, by means of Assumptions \ref{Rosenblatt_mix} and \ref{Markov_type} on the same sigma-fields $\mathcal{D}_t$. The verification of RC \ref{Condition_Cramer} is more technical and put in the online Supplementary Material. Here, we summarize the result in Lemma \ref{Taper_Cramer}:
\begin{lemma}\label{Taper_Cramer}
If RC \ref{Condition_Cramer} holds for  $\{g_t\}_{t=1}^{T}$, then it holds for $\{w(t)g_t\}_{t=1}^{T}$, with 
$w(t) = B_T^{-1/2}\sum_{s=1-t}^{T-t}k(s/B_T).$
\end{lemma}
Therefore, the validity of Assumptions \ref{a1}---\ref{Markov_type} and Lemma \ref{Taper_Cramer} imply that we can suitably approximate the probability distribution of $T^{1/2}\bar{g}_T$ by an Edgeworth expansion of the same kind as the one for $T^{1/2}\bar{g}$, when the latter exists. To derive the Edgeworth expansion for  $\hat{S}$, we have to adapt the expansion for $T^{1/2}\bar{g}_T$, as defined in \cite{gotze_asymptotic_1994}, to accommodate our studentization as in  \eqref{Eq S_psi}. To this end, we modify the proof of \cite{gotze_second-order_1996} to consider the tapering related to $\{w(t)\}_{t=1}^{T}$. 
To define our Edgeworth expansion, we need the following quantities:
\begin{align}
&\pi^\dag_T := T^{-1} \displaystyle\sum_{t=1}^{T} w(t) \displaystyle\sum_{j=1}^{T-B_T} \displaystyle\sum_{s=-B_T}^{B_T} k_T^{\ast}\left(s/B_T\right) \mathbb{E} \left[ g_t g_j g_ {j+s} \right],\label{cum1}
\\
&V_T:=\displaystyle\sum_{s=-B_T}^{B_T} k_T^{\ast}\left(s/B_T\right) T^{-1} \sum_{j=1}^{T-B_T}\left(g_{j}(\hat{\beta}) g_{j+s}(\hat{\beta})- \mathbb{E}\left[g_0 g_{s}\right]\right), 
\\
&\sigma_T^{\dag 2} := \displaystyle\sum_{s=-T}^{T}\left(1-\frac{\left\lvert s \right\rvert}{T}\right)w(s)\mathbb{E}\left[g_0 g_s \right],
\\
&\tau_{1T}^{2} := \displaystyle\sum_{s=-T}^{T} k_T^{\ast}\left(s/B_T\right)\mathbb{E}\left[g_0 g_{s} \right],
\\
&\mu^\dag_{3,T} := T^{2} \mathbb{E}\left[\bar{g}_T^3\right].\label{cum5}
\end{align}

\noindent We take the expansion $\hat{\sigma}^2 = \sigma_T^{\dag 2} + V_T + (\tau_{1T}^{2} - \sigma_T^{\dag 2}) + O_p(B_T/T)$, where $(\tau_{1T}^{2} - \sigma_T^{\dag 2})$ reflects the bias of the estimator $\hat{\sigma}.$
Then, we approximate $\hat{\sigma}^{-1}$ by the linear part of the Taylor series $\hat{\sigma}^{-1} = \sigma_T^{\dag -1} + \sum_{j=1}^{\infty} h^{(j)}(\sigma_T^{\dag 2}) (\hat{\sigma}^2-\sigma_T^{\dag 2})^j /j!$, where $h(x)=x^{-1/2}$ and $h^{(j)}$ is its $j$-th derivative, getting $\hat{\sigma}^{-1} = \sigma_T^{\dag -1} -(1/2)( V_T + (\tau_{1T}^{2} - \sigma_T^{\dag 2}) + O_p(B_T/T))\sigma_T^{\dag -3}.$ 
Taking the product with $T^{1/2}\bar{g}_T$, we obtain:

\begin{align}\label{ExpandS}
\hat{S} &= T^{1/2}\bar{g}_{T}\sigma_T^{\dag -1} - (1/2)T^{1/2}\bar{g}_{T}V_T \sigma_T^{\dag -3} - (1/2)T^{1/2}\bar{g}_{T}(\tau^2_{1T}-\sigma_T^{\dag 2})\sigma_T^{\dag -3} + O_p(B_T/T),
\\
&= E_T - (1/2)T^{1/2}\bar{g}_{T}(\tau^2_{1T}-\sigma_T^{\dag 2})\sigma_T^{\dag -3} + O_p(B_T/T),
\end{align}
\noindent where $E_T:=T^{1/2}\bar{g}_{T}\sigma_T^{\dag -1} - (1/2)T^{1/2} \bar{g}_{T} V_T \sigma_T^{\dag -3}$. Following the standard argument of \cite{chibisov_asymptotic_1972}, the Edgeworth expansion of $\hat{S}$ coincides with the one of $E_T$ up to the order $O(T^{1/2}\bar{g}_{T}(\tau^2_{1T}-\sigma_T^{\dag 2})\sigma_T^{\dag -3} + B_T/T)$. As $T^{1/2}\bar{g}_{T}$ is $O_p(1)$, the term $T^{1/2}\bar{g}_{T}(\tau^2_{1T}-\sigma_T^{\dag 2})\sigma_T^{\dag -3}$ is of order $O_p(\lvert \tau^2_{1T}-\sigma_T^{\dag 2} \rvert) = O_p\left(B_T^{-q}\right)$, and corresponds to the bias of the variance estimator.

To assess the order of the FMB approximation error, it is more convenient to work directly with the Fourier transform of the Edgeworth expansion; see \cite{gotze_second-order_1996}, p.\ 1919. Therefore, we define the Edgeworth expansion of $\hat{S}$, say $\Upsilon^{\dag}_{S,T},$ in terms of its Fourier transform. We do so by collecting the expansion of $E_T$ and the error terms discussed above:
\begin{align}
\mathcal{E}^{\dag}_T \left( \tau \right) &:= \int \exp \left( i\tau^\intercal x \right) d\Upsilon^{\dag}_{S,T}\left( x \right)\nonumber
\\
&= \left\{ 1 + \frac{1}{\sqrt{T}} \frac{1}{\sigma^{\dag3}_T} \left[ \left( \frac{\mu^{\dag}_{3,T}}{6} - \frac{\pi^{\dag}_T}{2} \right) \left( i\tau \right)^3 - \frac{\left( i\tau \right) \pi^{\dag}_T}{2} \right] + O(B_T^{-q}) + O(B_T/T)\right\} \exp \left( -\frac{\tau^2}{2} \right).\label{Edg0}
\end{align}
Invoking Esseen Lemma, we get the approximation error of $\Upsilon^{\dag}_{S,T}$ as in Theorem \ref{Ed_Bound0}, Equation \eqref{EE1}. \\

We provide also a sketch for the derivation of Edgeworth expansion for $\hat{Q}$. The detailed calculation is available in the Supplementary Material. To get an Edgeworth expansion for $\hat{Q}$, namely  \eqref{EE2} in Theorem \ref{Ed_Bound0}, we make use of the univariate Edgeworth expansion $\Upsilon^{\dag}_{S,T}$, explicitly defined in \eqref{EE1}. 

First, when $T^{1/2}\bar{g}_{T}$ is multivariate of dimension $r$, expansions of the same form as in Theorem \ref{Ed_Bound0} Equation \eqref{EE1} hold (up to constants independent of $T$) for any linear combination $T^{1/2} \upsilon ^{\intercal} \bar{g}_{T} / (\upsilon ^{\intercal}\hat{\Omega} \upsilon)^{1/2}$ (\cite{gotze_second-order_1996}). 

Second, as $\hat{\Omega}$ is symmetric positive semi-definite by construction, we have a unique symmetric positive semi-definite square root $\hat{\Omega}^{1/2}$, which admits an inverse. Thus, we have a vector $\hat{Q}^{1/2}:=\hat{\Omega}^{-1/2}T^{1/2}\bar{g}_{T},$ such that $\hat{Q}^{1/2 \intercal}\hat{Q}^{1/2} = \hat{Q}.$ Projecting the vector $T^{1/2}\bar{g}_{T}$ onto the orthonormal eigenvectors of $\hat{\Omega}$, we get $\hat{Q}^{1/2} = \Lambda^{-1/2}P^{\intercal}T^{1/2}\bar{g}_{T} = (T^{1/2}\bar{g}_{T}^{\intercal}v_1/\lambda_1^{1/2},...,T^{1/2}\bar{g}_{T}^{\intercal}v_r/\lambda_r^{1/2} )^{\intercal}$, where $\{\lambda_1,...,\lambda_r\}$ are the eigenvalues of $\hat{\Omega}$ corresponding to its normalized eigenvectors $\{v_1,...,v_r\}$, $\Lambda := \diag(\lambda_1,...,\lambda_r)$, and $P:=(v_1,...,v_r)$. As $T^{1/2}\bar{g}_{T}^{\intercal}v_j/\lambda_j^{1/2} = T^{1/2} \upsilon ^{\intercal} \bar{g}_{T} / (\upsilon ^{\intercal}\hat{\Omega} \upsilon)$ when we choose $\upsilon = v_j$ for each $j=1,...,r$, we directly see that there exist expansions of the same form as in Theorem \ref{Ed_Bound0} Equation \eqref{EE1} for each element of $\hat{Q}^{1/2}$. Furthermore, taking any vector $c$ such that $\|c\|=1$, there exists a univariate expansion of the same form as in Theorem \ref{Ed_Bound0} Equation \eqref{EE1} for $c ^{\intercal} \hat{Q}^{1/2} / (c ^{\intercal}I_r c)^{1/2} = c ^{\intercal} \hat{Q}^{1/2}$, as the variance estimator of $\hat{Q}^{1/2}$ is the $r$-dimensional identity matrix $I_r$ by definition. By the Cramér-Wold device, the characteristic function of $\hat{Q}^{1/2}$ is $\mathcal{E}_{r}(\tau c):=\mathbb{E}[\exp ( i \tau c^\intercal \hat{Q}^{1/2})]$, where $\tau$ is a scalar. As a consequence, there exists an expansion of the same form as in \eqref{Edg0} for $\mathcal{E}_{r}(\tau c)$. Taking the inverse Fourier transform of this approximation, we approximate the probability distribution of $\hat{Q}^{1/2}$ by a multivariate Edgeworth expansion $\Upsilon _ {r , T } ( z ) : = \Phi_r(z)+T^{-1/2}p_1(z)\phi_r(z)+(B_T/T)p_2(z)\phi_r(z)$, where $\phi_r(z)$ is the normal density on $\mathbb{R}^r$ with mean zero and variance $I_r$ and $\Phi_r(z)$ is the corresponding cdf. In this expansion, $p_1$ is en even polynomial in $z$ and $p_2$ is an odd polynomial in $z$. The error of this approximation is by construction the same as the univariate one, as it is built with the same type of variance estimator.

Finally, we consider the statistic of interest $\hat{Q}$ and we work on a variable transform. Indeed, 
$$\sup_{x \in \mathbb{R}^r} | \int_{t \leq x} d\Upsilon_{r,T} - \int_{t \leq y} dF_{\hat{Q}^{1/2}}| = o(B_T/T) + O( B_T^{-q} ) + O(B_T/T)$$ 
implies $\sup_{y \in \mathbb{R}_{+}} | \int_{\{t : t^{\intercal} t \leq y\}} d\Upsilon_{r,T} - \int_{t \leq y} dF_{\hat{Q}}| =  o(B_T/T) + O( B_T^{-q} ) + O(B_T/T)$. Along the same lines of \cite{chandra_valid_1979} (see Supplementary Material), we identify an expansion $\Upsilon^{\dag}_{Q,T}$ such that $\sup_{y \in \mathbb{R}_+} | \int_{t \leq y} d\Upsilon^{\dag}_{Q,T} - \int_{t \leq y} dF_{\hat{Q}}| =  o(B_T/T) + O( B_T^{-q} ) + O(B_T/T)$. In this expansion, the elementary probability measure is $\mathcal{X}^2_r$ instead of the Gaussian $\Phi_r$, and the term of order $T^{-1/2}$ disappears because $p_1$ is even in $z$. 
In principle, we can find explicitly $p_1$ by inverting the Fourier transform in \eqref{Edg0}. However, we do not need $p_1$ in this proof since we work directly with the Fourier transform. Likewise, we only give the polynomials $p_2$ and $p_Q$ formally (implicitly), as we do not need their explicit form in the sequel of the proof. What matters is the order of error to which they correspond, namely $O(B_T/T)$.

\subsection*{A.2. Edgeworth expansion of the bootstrap sample statistics}\label{Bootstrap Sample Edgeworth}

In the sequel, $\mathbb{E}^{\ast}$ denotes the expectation under the bootstrap probability measure as in 
\eqref{boot_distrib}. We state the following:
\begin{theorem}\label{Ed_Bound1}
Under Assumptions \ref{a1}, \ref{a2}, \ref{EE5} and if $\mathbb{E} \left[ \left\| g_t \right\|^{\bar q s + \delta} \right] < \infty$, for $\delta > 0$, $s \geq 8$, and $\bar q \geq 3$:
\begin{align}
\Upsilon^{\ast}_{S,T}( x ) &= \Phi ( x ) + T ^ { - 1 / 2 } p _ { 1 } ( x,\mathcal{K}_S^{1 \ast}) \phi ( x ) + T ^ { - 1 } p _ { 2 } ( x,\mathcal{K}_S^{2 \ast}) \phi ( x ),\label{BootEdgS}
\\
\Upsilon^{\ast}_{Q,T} ( x ) &= F _ { \mathcal{X} _ { r } ^ { 2 } } ( x ) + T ^ { - 1 } p _ { Q } ( x,\mathcal{K}_Q^{\ast} ) f _ { \mathcal{X} _ { r } ^ { 2 } } ( x ),\label{BootEdgQ}
\end{align}
with uniform error bound:
\begin{align}
\sup_{x \in \mathbb{R}}\left\lvert \mathbb{P}^{\ast}\left[S^{\ast}\leq x\right] - \Upsilon^{\ast}_{S,T} ( x ) \right\rvert =& o_p\left(T^{-1}\right), \label{UEB3}
\\
\sup_{x \in \mathbb{R}_{+}}\left\lvert \mathbb{P}^{\ast}\left[ Q^{\ast} \leq x\right] - \Upsilon^{\ast}_{Q,T} ( x ) \right\rvert =& o_p\left(T^{-1}\right),
\end{align}
where $\mathcal{K}_S^{1 \ast}$, $\mathcal{K}_S^{2 \ast}$ and $\mathcal{K}_Q^{\ast}$ are the bootstrap counterparts of the vectors $\mathcal{K}_S^1$, $\mathcal{K}_S^2$ and $\mathcal{K}_Q$. \label{UEB4}
\end{theorem}

\textbf{Proof of Theorem \ref{Ed_Bound1}}. To prove the theorem, we need to define the Edgeworth expansion for the bootstrap statistic. To this end, we need the same arguments as for the original statistic. Namely, we need to check RC \ref{a1}---\ref{Markov_type} for the bootstrap sample $\{ g_{T,t}^{\ast} \}_{t=1}^{T}$, conditionally on $\left\{ g_{t} \left( \beta_0 \right) \right\}_{t=1}^{T}$, uniformly on a set whose probability tends to one. When $p=r$ (exactly-identified moment conditions), RC \ref{a1} holds by the nature of the estimating equations, as $\mathbb{E}^{\ast} \left[ g_{T,t}^{\ast} \right]= \bar{g}_T \left( \hat{\beta} \right) = 0$. In the case of over-identification, we have to recenter the bootstrap statistic so that RC \ref{a1} holds. For RC \ref{a2}, we have $\mathbb{E}^{\ast} \left[ \left\| g_{T,t}^{\ast} \right\|^s \right] = T^{-1} \displaystyle\sum_{t=1}^{T} \left\| g_{T,t} \right\|^s$. For the unconditional moments, we get $\mathbb{E}\left[ \mathbb{E}^{\ast} \left[ \left\| g_{T,t}^{\ast} \right\|^s \right] \right] = T^{-1}\sum_{t=1}^{T}\mathbb{E} \left[ \left\| g_{T,t} \right\|^s \right] = \mathbb{E} \left[ \left\| g_{T,t} \right\|^s \right] < \infty$, from Assumption \ref{a2} on $g_t$ and by Minkowski inequality. Then, assuming $\mathbb{E} \left[ \left\| g_t \right\|^{\bar qs} \right] < \infty$, for $\bar q \geq 3$, we obtain $\mathbb{E}^{\ast} \left[ \left\| g_{T,t}^{\ast} \right\|^s \right] - \mathbb{E}\left[ \mathbb{E}^{\ast} \left[ \left\| g_{T,t}^{\ast} \right\|^s \right] \right] = O_p(T^{-1/2})$. As a consequence, $\mathbb{E}^{\ast} \left[ \left\| g_{T,t}^{\ast} \right\|^s \right]$ is bounded with probability tending to one.

RC \ref{RV_approx}, \ref{Rosenblatt_mix}, and \ref{Markov_type}, trivially hold by independence of the $g_{T,t}^{\ast}$ under the bootstrap probability measure as in \eqref{boot_distrib}, when $\mathcal{D}_t := \sigma\langle g_{T,t}^{\ast} \rangle$ for $t=1,...,T$. Finally, as the bootstrap random variables are i.i.d, we can show that the standard Cramér condition holds asymptotically, instead of checking RC \ref{Condition_Cramer}. Thus, the Cramér condition in Assumption \ref{EE5} is sufficient to verify RC \ref{Condition_Cramer} for the bootstrap process, with the same sub-sigma-fields $\mathcal{D}_t.$ Indeed, from Assumption \ref{EE5}, we have : $$f(\beta_0)=\limsup_{T \rightarrow \infty} \sup_{b < \lvert \tau \rvert < e^{\delta T}} \lvert T^{-1}\sum_{t=1}^{T} \exp \left( i  \tau g_{T,t}\left(\beta_0\right) \right) \rvert < 1, \text{ } a.s.$$ We have to make sure that the same limit holds when $g_{T,t}(\beta)$ is evaluated at $\hat{\beta}$ instead of $\beta_0$, since we resample the $\{ g_{T,t}(\hat{\beta}) \}_{t=1}^{T}$ in FMB. From Assumption \ref{EE5}, there exists a constant $c>0$ such that $\sup_{\left\| \beta - \beta_0 \right\| \leq c} f(\beta) < 1.$ As $\lim_{T \rightarrow \infty}\| \hat{\beta} - \beta_0 \| \leq c,$ we have a.s. $f(\hat \beta) \leq \sup_{\left\| \beta - \beta_0 \right\| \leq c} f(\beta) < 1,$ which verifies the necessary Cramér condition.

We construct the Edgeworth expansion $\Upsilon^{\ast}_{S,T}$ of the bootstrap statistic $S^{\ast}$ in the same way as the one for the original sample statistic $\hat{S}$ in \eqref{Edg0} above. Thus, we need to define the bootstrap counterpart of the quantities \eqref{cum1}---\eqref{cum5}:
\begin{align}
&\pi^{\ast}_T := T^{-1} \sum_{t=1}^{T} \sum_{j=1}^{T} \sum_{s= 1}^{T} \mathbb{E}^{\ast} [ g^{\ast}_{T,t} g^{\ast}_{T,j} g^{\ast}_{T,s}],
\\
&\sigma^{\ast 2}_T := \mathbb{E}^{\ast} \left[ (T^{1/2}\bar{g}_{T}^{\ast})^2 \right],
\\
&\tau_{1T}^{\ast 2} := \displaystyle\sum_{s=-T}^{T} \mathbb{E}^{\ast}\left[g_0^{\ast}g_s^{\ast}\right],
\\
&\mu^{\ast}_{3,T} := T^{1/2} \mathbb{E}^{\ast} \left[ (T^{1/2}\bar{g}_{T}^{\ast})^3 \right].
\end{align}
Then, we can define the bootstrap Edgeworth expansion {via} its Fourier transform:
\begin{align}
\mathcal{E}^{\ast}_T \left( \tau \right) &:= \int \exp \left( i\tau^\intercal x \right) d\Upsilon^{\ast}_{S,T}\left( x \right) \nonumber
\\
&= \left\{ 1 + \frac{1}{\sqrt{T}} \frac{1}{\sigma^{\ast3}_T} \left[ \left( \frac{\mu^{\ast}_{3,T}}{6} - \frac{\pi^{\ast}_T}{2} \right) \left( i\tau \right)^3 - \frac{\left( i\tau \right) \pi^{\ast}_T}{2} \right] + O_p(\lvert \sigma^{\ast 2}_T - \tau^{\ast 2}_{1T} \rvert) + O_p(1/T)\right\} \exp \left( -\frac{\tau^2}{2} \right). \nonumber
\end{align}
We have the same error bound as in Theorem \ref{Ed_Bound0} up to four aspects. First, $\Upsilon^{\ast}_{S,T}$ is a random measure because it is conditional to the original sample. Second, by independence of the bootstrap variables, we have $\sigma_T^{\ast 2} = \tau_{1T}^{\ast 2},$ so the contribution of the $\lvert \sigma^{\ast 2}_T - \tau^{\ast 2}_{1T} \rvert$ term to the error disappears. Third, in contrary to \eqref{Edg0}, we have the order $O_p(1/T)$ instead of $O_p(B_T/T),$ as it is an Edgeworth expansion for the studentized mean of i.i.d.\ random variables. Finally, by independence of the $\{g^{\ast}_{T,t}\},$ we have $\pi^{\ast}_T = T^{-1} \sum_{t=1}^{T} \sum_{j=1}^{T} \sum_{s=1}^{T} \mathbb{E}^{\ast} [ g^{\ast}_{T,t} g^{\ast}_{T,j} g^{\ast}_{T,s}] = T^2 \mathbb{E}^{\ast} \left[ \bar{g}_{T}^{\ast 3} \right] = \mu^{\ast}_{3,T}.$
Altogether, it leads to:
\begin{align}
\mathcal{E}^{\ast}_T \left( \tau \right) &= \left\{ 1 + \frac{1}{\sqrt{T}} \frac{1}{\sigma^{\ast3}_T} \left[ \left( \frac{\mu^{\ast}_{3,T}}{6} - \frac{\mu^{\ast}_{3,T}}{2} \right) \left( i\tau \right)^3 - \frac{\left( i\tau \right) \mu^{\ast}_{3,T}}{2} \right] + O_p(1/T)\right\} \exp \left( -\frac{\tau^2}{2} \right). \nonumber
\\
&= \left\{ 1 + \frac{\mu^{\ast}_{3,T}}{\sqrt{T} \sigma^{\ast 3}_T} \left[ -\frac{\left( i\tau \right)^3}{3} - \frac{\left( i\tau \right)}{2} \right] + O_p(1/T) \right\} \exp \left( -\frac{\tau^2}{2} \right). \label{Edg1}
\end{align}
To derive an Edgeworth expansion for $Q^{\ast}$, we proceed along the same lines as for the original sample statistic $\hat{Q}$, working with univariate Edgeworth expansions for studentized linear combinations of the form $T^{1/2} \upsilon^{\intercal} \bar{g}^{\ast}_{T}$.

\subsection*{A.3. Higher-order correctness of FMB}

To prove part (i) of Theorem \ref{higher_order}, we essentially need the convergence of the terms of order $T^{-1/2}$ in $\Upsilon^{\dag}_{S,T}$ (as in \eqref{EE1}) and $\Upsilon^{\ast}_{S,T}$ (as in \eqref{BootEdgS}) to the same quantities. Indeed, in view of Theorems \ref{Ed_Bound0} and \ref{Ed_Bound1}, the condition $\plim_{T \rightarrow \infty} \sup_{x \in \mathbb{R}} \lvert p_1(x,\mathcal{K}_S^{1 \ast}) - p_1(x,\mathcal{K}_S^1) \rvert = 0$, or equivalently $\mathcal{K}_S^{1 \ast} = \mathcal{K}_S^1 + o_p(1),$ is sufficient to show higher-order refinements of FMB. We give the details of this convergence in the proof of Lemma \ref{Fourier_Edge_Bound} in the online Supplementary Material.
\begin{lemma}\label{Fourier_Edge_Bound}
Under Assumptions \ref{a1}---\ref{EE5} and if $\mathbb{E} \left[ \left\| g_t \right\|^{\bar qs + \delta} \right] < \infty$, for $\delta > 0$, $s \geq 8$, and $\bar q \geq 3$:
$$\sup_{x \in \mathbb{R}} \lvert \Upsilon^{\dag}_{S,T} \left( x \right) - \Upsilon^{\ast}_{S,T} \left( x \right) \rvert = o_p(T^{-1/2}) + O_p\left(B_T^{-q}\right) + O_p(B_T/T).$$
\end{lemma}
Collecting the error bounds of Theorem \ref{Ed_Bound0}, Theorem \ref{Ed_Bound1}, and Lemma \ref{Fourier_Edge_Bound}, we use the triangular inequality to get:
\begin{align*}
\sup_{x \in \mathbb{R}}\left\lvert \mathbb{P}^\ast\left[S^{\ast} \leq x\right] - \mathbb{P}\left[\hat{S}_{} \leq x\right]\right\rvert &\leq \sup_{x \in \mathbb{R}}\left\lvert \mathbb{P}^{\ast}\left[S^{\ast}\leq x\right] - \Upsilon^{\ast}_{S,T} ( x ) \right\rvert + \sup_{x \in \mathbb{R}} \lvert \Upsilon^{\ast}_{S,T}\left( x \right) - \Upsilon^{\dag}_{S,T} \left( x \right) \rvert
\\
&\qquad + \sup_{x \in \mathbb{R}}\left\lvert \Upsilon^{\dag}_{S,T} ( x ) - \mathbb{P}\left[\hat{S}_{} \leq x\right] \right\rvert,
\\
&= o_p\left(T^{-1/2}\right) + O_p\left(B_T^{-q}\right) + O_p(B_T/T).
\end{align*}

The proof of part (ii) of Theorem \ref{higher_order} is more direct, as the even polynomial $p_1$ in Theorems \ref{Ed_Bound0} and \ref{Ed_Bound1} deletes the $T^{-1/2}$ term from the Edgeworth expansion of $\hat{Q}$. Hence, only the higher-order terms remain. As a consequence, we immediately verify that $\sup_{x \in \mathbb{R^{+}}} \lvert \Upsilon^{\ast}_{Q,T}\left( x \right) - \Upsilon^{\dag}_{Q,T} \left( x \right) \rvert$ is of order $O_p\left(B_T^{-q}\right) + O_p(B_T/T).$ Then, we prove part (ii) making use of Theorems \ref{Ed_Bound0}, \ref{Ed_Bound1}, and the triangular inequality:
\begin{align*}
\sup_{x \in \mathbb{R}^{+}} \left\lvert \mathbb{P}^\ast\left[ Q^{\ast} \leq x\right] - \mathbb{P}\left[\hat{Q} \leq x\right]\right\rvert &\leq \sup_{x \in \mathbb{R}^{+}} \left\lvert \mathbb{P}^{\ast}\left[Q^{\ast}\leq x\right] - \Upsilon^{\ast}_{Q,T} ( x ) \right\rvert + \sup_{x \in \mathbb{R}^{+}} \lvert \Upsilon^{\ast}_{Q,T}\left( x \right) - \Upsilon^{\dag}_{Q,T} \left( x \right) \rvert
\\
& \qquad  + \sup_{x \in \mathbb{R}^{+}} \left\lvert \Upsilon^{\dag}_{Q,T} ( x ) - \mathbb{P}\left[\hat{Q}_{} \leq x\right] \right\rvert,
\\
&= O_p\left(B_T^{-q}\right) + O_p(B_T/T).
\end{align*}

\newpage
\section*{}
\begin{center} 
\Large{\textbf{A Higher-Order Correct Fast Moving-Average Bootstrap \\ for Dependent Data}} \Large{\\
\vspace{0.5cm}
Online Supplementary Material} \\
\vspace{0.25cm}
\normalsize{Davide La Vecchia, Alban Moor and Olivier Scaillet}
\end{center}
\vspace{0.5cm}
\hrule
\vspace{0.5cm}
This Supplementary Material contains the proofs of Lemma \ref{Taper_Cramer} and Lemma \ref{Fourier_Edge_Bound}, whose proof requires two additional Lemmas. We provide also 
the key points of the derivation of the Edgeworth expansion for the quadratic statistics $\hat{Q}$ and $Q^{\ast}$, as well as the proof of Corollary \ref{HOtildeQ}. In the sequel of this Supplementary Material, we provide the detailed comparison of CPU time between FMB and MBB (Table \ref{Tab: CPU}), a complement of Tables \ref{Tab: 10_025COVER_3} and \ref{Tab: 025COVER_3} with a smaller sample size $T=150$ in Table \ref{Tab: T150} and a graphical illustration of the use of $\tilde{Q}$ as in Remark \ref{nomono}.
\vspace{0.5cm}
\hrule
\vspace{1cm}

\subsection*{SM.1. Proof of Lemma \ref{Taper_Cramer}}

Without loss of generality, take $D_t:= \sigma\langle g_{t,0}^\ddagger\rangle$. Then, let us define $\mathcal{F}_m$ as the set of index $k \in \{-m,...,m\}$ such that $g_{t+k}$ is $\{\mathcal{D}_j : j \neq t\}$-measurable. We can now rewrite RC \ref{Condition_Cramer} as follows:
\begin{align*}
\exp(-\delta) &\geq \mathbb{E}[\lvert\mathbb{E}[\exp(i\tau ^{\intercal} (g_{t-m} + ... + g_{t+m}) ) | \mathcal{D}_j : j \neq t]\rvert],
\\
&=\mathbb{E}[\lvert\mathbb{E}[\exp(i\tau ^{\intercal} \sum_{k \in \mathcal{F}_m} g_{t+k} ) \exp(i\tau ^{\intercal} \sum_{\ell \in \mathcal{F}_m^{\complement}} g_{t+\ell} ) | \mathcal{D}_j : j \neq t]\rvert],
\\
&=\mathbb{E}[\lvert\exp(i\tau ^{\intercal} \sum_{k \in \mathcal{F}_m} g_{t+k} ) \mathbb{E}[\exp(i\tau ^{\intercal} \sum_{\ell \in \mathcal{F}_m^{\complement}} g_{t+\ell}) | \mathcal{D}_j : j \neq t ]\rvert],
\\
&=\lvert\mathbb{E}[\exp(i\tau ^{\intercal} \sum_{\ell \in \mathcal{F}_m^{\complement}} g_{t+\ell} ) | \mathcal{D}_j : j \neq t ]\rvert \mathbb{E}[\lvert\exp(i\tau ^{\intercal} \sum_{k \in \mathcal{F}_m} g_{t+k})\rvert],
\\
&=\lvert\mathbb{E}[\exp(i\tau ^{\intercal} \sum_{\ell \in \mathcal{F}_m^{\complement}} g_{t+\ell} ) | \mathcal{D}_j : j \neq t ] \rvert.
\end{align*}
Therefore, we have to show that $\lvert\mathbb{E}[\exp(i\tau ^{\intercal} \sum_{\ell \in \mathcal{F}_m^{\complement}} g_{t+\ell} ) | \mathcal{D}_j : j \neq t ] \rvert \leq \exp(-\delta)$ implies $\lvert\mathbb{E}[\exp(i\tau ^{\intercal} \sum_{\ell \in \mathcal{F}_m^{\complement}} w(t+\ell)g_{t+\ell} ) | \mathcal{D}_j : j \neq t ] \rvert \leq \exp(-\delta)$. First, the expectation is taken with respect to the same conditional probability measure, say $F(g_{t+\ell} : \ell \in \mathcal{F}_m^{\complement})$ for convenience of notation. Then, $\lvert\mathbb{E}[\exp(i\tau ^{\intercal} \sum_{\ell \in \mathcal{F}_m^{\complement}} w(t+\ell)g_{t+\ell} ) | \mathcal{D}_j : j \neq t ] \rvert = \lvert \int \exp(i\tau ^{\intercal} \sum_{\ell \in \mathcal{F}_m^{\complement}} w(t+\ell)g_{t+\ell} ) dF(g_{t+\ell} : \ell \in \mathcal{F}_m^{\complement}) \rvert$. It is now clear that the weights $w(t+\ell)$ are rescaling horizontally the same characteristic function. As $w(t)\neq 0,$ $\forall t$, the complex modulus of the characteristic function never reaches unity (corresponding to the infimum $\delta=0$). Thus, the range of $\mathbb{E}[\exp(i\tau ^{\intercal} \sum_{\ell \in \mathcal{F}_m^{\complement}} g_{t+\ell} ) | \mathcal{D}_j : j \neq t ]$ is the same than the range of $\mathbb{E}[\exp(i\tau ^{\intercal} \sum_{\ell \in \mathcal{F}_m^{\complement}} w(t+\ell)g_{t+\ell} ) | \mathcal{D}_j : j \neq t ]$. As a consequence, the upper bound of their complex modulus is the same, $\forall \| \tau \| > 0$.

For the second part of the condition, it is immediate that $\liminf_{T \rightarrow \infty} T^{-1} \Var \left[ \sum_{t=1}^{T} w(t)g_t \right] > 0,$ if the same condition holds for $\{ g_t \}_{t=1}^{T}$, as $w(t) \neq 0$, $\forall t$.

\subsection*{SM.2. Edgeworth expansion for the quadratic statistics $\hat{Q}$ and $Q^{\ast}$}

For the completeness of the presentation, we supplement here the arguments of the paragraph before Theorem \ref{Ed_Bound0}, following closely the lines of \cite{chandra_valid_1979} with their notation. The changes concern the order of approximations in the Edgeworth expansion and the specification of the statistic of interest.

First, consider the Edgeworth expansion $\Upsilon _ {r , T } ( z ) : = \Phi_r(z)+T^{-1/2}p_1(z)\phi_r(z)+(B_T/T)p_2(z)\phi_r(z),$
which approximates the distribution of the $r$-dimensional statistic 
$\hat{Q}^{1/2} = \Lambda^{-1/2}P^{\intercal}T^{1/2}\bar{g}_{T} = (T^{1/2}\bar{g}_{T}^{\intercal}v_1/\lambda_1^{1/2}, \\ ...,T^{1/2}\bar{g}_{T}^{\intercal}v_r/\lambda_r^{1/2} )^{\intercal},$ 
with an error of order $o(B_T/T) + O(| \tau^{2}_{1T} - \sigma^{\dag 2}_T |)$. For ease of notation, let us take $Z:=\hat{Q}^{1/2}$. In the paragraph before Theorem \ref{Ed_Bound0}, we obtain the bound $\sup_{y \in \mathbb{R}_{+}} | \int_{\{t : t^{\intercal} t \leq y\}} d\Upsilon_{r,T} - \int_{t \leq y} dF_{\hat{Q}}| = o(B_T/T) + O( | \tau^{2}_{1T} - \sigma^{\dag 2}_T | )$. Thus, we need to identify an expansion of the form $\Upsilon^{\dag}_{Q,T} ( x ) = F _ { \mathcal{X} _ { r } ^ { 2 } } ( x ) + (B_T / T ) p _ { Q } ( x,\mathcal{K}_Q) f _ { \mathcal{X} _ { r } ^ { 2 } } ( x )$ such that $\sup_{y \in \mathbb{R}_+} | \int_{t \leq y} d\Upsilon^{\dag}_{Q,T} - \int_{t \leq y} dF_{\hat{Q}}| = o(B_T/T) + O( | \tau^{2}_{1T} - \sigma^{\dag 2}_T |)$. 

For $s=5$, our $(s-3)$-order Edgeworth expansion $\Upsilon_{r,T}(z)$ has a density of the form $\tilde{\upsilon}_{r,T}(z) := [1+T^{-1/2}\tilde{p}_1(z)+(B_T/T)\tilde{p}_2(z)]\phi_r(z)$, where the degree of each term in the polynomials $\tilde{p}_{1}(z)$ and $\tilde{p}_{2}(z)$ is respectively odd and even. 

Now, we explain how to obtain the expansion $\Upsilon^{\dag}_{Q,T}(v)$, with density of the form $\tilde{\upsilon}^{\dag}_{Q,T}(v) : = [1 + (B_T / T ) \tilde{p} _ { Q } (v) ]f _ { \mathcal{X} _ { r } ^ { 2 } } ( v )$. This density is similar to $\tilde{\upsilon}_{r,T}(z)$, except for the $\mathcal{X} _ { r }^2$ measure replacing the Gaussian $\Phi_r$ measure, and the cancellation of the term of order $T^{-1/2}$.

Consider the multivariate polar transformation $T_1$, which sends $Z$ to $(R,\mathbf{\beta})^{\intercal}:=(R,\beta^{(1)},...,\beta^{(r-1)})^{\intercal}$ via $z_1 = R \prod_{i=1}^{r-1} \cos(\beta^{(i)})$ and $z_j = R \sin(\beta^{(r-j+1)})\prod_{i=1}^{r-j} \cos(\beta^{(i)})$, where $2 \leq j \leq r$, $R \in \mathbb{R}_{+}^{\ast}$ is the radius, and $-\pi/2 < \beta^{(i)} < \pi/2$ for $i=1,...,r-2$, $0 \leq \beta^{(r-1)} \leq 2\pi$ are the angles. Then, for a vector of non-negative integers $A := (a_1,...,a_r)$, we write $\mathcal{R}(A) = \mathcal{R}(R,\mathbf{\beta},Z) = R^{a_0}\prod_{i=1}^{r}(z_i/R)^{a_i} = \prod_{i=1}^{r} z_i^{a_i}$, where $a_0=\sum_{i=1}^{r}a_i$. We will use notation $\mathcal{R}(A)$ even when $a_0 \neq \sum_{i=1}^{r}a_i$. We say that $\mathcal{R}(A)$ is odd if at least one element of $A$ is odd, and more generally, we say that the expression $R^{a_0}\prod_{i=1}^{r-1}\cos(\beta^{(i)})^{a_i}\sin(\beta^{(i)})^{b_i}$ is odd if at least one of $\{b_1,...,b_{r-1},a_{r-1}\}$ is odd. The Jacobian of $T_1$ is $R^{r-1}\prod_{i=1}^{r-2}\cos(\beta^{(i)})^{r-i-1}$, say $R^{r-1}J(\mathbf{\beta})$ for ease of notation, and that an odd $\mathcal{R}(A)$ implies an odd $\mathcal{R}(A)J(\mathbf{\beta})$. Finally, we write $R_{i,j}(R,\mathbf{\beta},Z)$ a finite sum of constant multiples of terms of the form $\mathcal{R}(A)$, and say that $R_{i,j}(R,\mathbf{\beta},Z)$ is odd if every such $\mathcal{R}(A)$ are odd. In the following proof, various expressions of the form $R_{i,j}(R,\mathbf{\beta},Z)$ occur with the two following properties. First, $R_{i,j}(R,\mathbf{\beta},Z)$ happens to be odd when $j$ is odd. Second, if $j$ is even and $R_{i,j}(R,\mathbf{\beta},Z)$ includes some (constant multiple) of $\mathcal{R}(A)$ which fails to be odd, then the power $a_0$ in the corresponding $\mathcal{R}(A)$ will be even.
Let us define the set $M_T :=\{ Z: \|Z\|^2 < (s-1) \log{T} \}$,
and for any $\bar{B} \subset \mathbb{R}$, write $\bar{B}_T := \{ Z: Z^{\intercal}Z/2 \in \bar{B} \}$.
It is sufficient to exhibit an expansion $\tilde{\upsilon}^{\dag}_{Q,T}(v)$ such that $\sup \{ | \int_{ \{\bar{B}_T \cap M_T \}} \tilde{\upsilon}_{r,T} - \int_{\{\bar{B}\}} \tilde{\upsilon}^{\dag}_{Q,T} | : \bar{B} \in \mathcal{B} \} = o(B_T/T) + O( | \tau^{2}_{1T} - \sigma^{\dag 2}_T |)$, where $\mathcal{B}$ are the Borel sets on $\mathbb{R}$. Applying $T_1$, we get $$\displaystyle\int\limits_{ \{\bar{B}_T \cap M_T \}} \tilde{\upsilon}_{r,T} = \displaystyle\int\limits_{ T_1(\bar{B}_T \cap M_T )} R^{r-1}J(\mathbf{\beta})[1+T^{-1/2}R_{1,1}(R,\mathbf{\beta},Z)+(B_T/T)R_{1,2}(R,\mathbf{\beta},Z)]\exp(-R^2/2).$$
\noindent Then, let us apply the transformation $T_2(R,\mathbf{\beta},Z) := (R^{\prime},\mathbf{\beta},Z)$, where $R^{\prime} := (T_1^{-1}(R,\mathbf{\beta},Z)^{\intercal}T_1^{-1}(R,\mathbf{\beta},Z)/2)^{1/2}$.
It can be shown that $R=R^{\prime}(1+ T^{-1/2}R_{3,1}(R,\mathbf{\beta},Z)+(B_T/T)R_{3,2}(R,\mathbf{\beta},Z) + o(B_T/T) + O( | \tau^{2}_{1T} - \sigma^{\dag 2}_T |))$ uniformly on $T_2 T_1(M_T)$.
The derivative of $\hat{Q}$ with respect to the radius $R$ is 
$$\partial (T_1^{-1}(R,\mathbf{\beta},Z)^{\intercal}T_1^{-1}(R,\mathbf{\beta},Z)/2) / \partial R = 2R (1+ T^{-1/2}R_{4,1}(R,\mathbf{\beta},Z)+(B_T/T)R_{4,2}(R,\mathbf{\beta},Z),$$ and the Jacobian of $T_2$ is $ \partial R / \partial R^{\prime} = 2R^{\prime}(\partial (T_1^{-1}(R,\mathbf{\beta},Z)^{\intercal} T_1^{-1}(R,\mathbf{\beta},Z)/2) / \partial R)^{-1}$. As a consequence, we obtain: 
\begin{align*}
\begin{split}
\displaystyle\int\limits_{ \{\bar{B}_T \cap M_T \}} \tilde{\upsilon}_{r,T} &= \displaystyle\int\limits_{\{(R,\mathbf{\beta},Z):(R^{\prime})^2 \in \bar{B}\} \cap T_2 T_1 (M_T)} \left[ (R^{\prime})^{r-1}J(\mathbf{\beta})[1+T^{-1/2}R_{5,1}(R^{\prime},\mathbf{\beta},Z) \right. \\ 
& \left. +(B_T/T)R_{5,2}(R^{\prime},\mathbf{\beta},Z)]\exp(-(R^{\prime})^2/2) \vphantom{T^{-1/2}} \right] + o(B_T/T) + O( | \tau^{2}_{1T} - \sigma^{\dag 2}_T |),
\end{split}
\end{align*}
\noindent uniformly on the Borel subsets on $\mathbb{R}$. Finally, as $ T_2 T_1 (M_T) \subset \{ (R^{\prime},\mathbf{\beta},Z):(R^{\prime})^2 < ((s-3)/2)\log T \} $, we have:
\begin{align*}
\begin{split}
\displaystyle\int\limits_{ \{\bar{B}_T \cap M_T \}} \tilde{\upsilon}_{r,T} &= \displaystyle\int\limits_{\{(R^{\prime},\mathbf{\beta},Z):(R^{\prime})^2 \in \bar{B}\}} \left[ (R^{\prime})^{r-1}J(\mathbf{\beta})[1+T^{-1/2}R_{5,1}(R^{\prime},\mathbf{\beta},Z) \right. \\ 
& \left. +(B_T/T)R_{5,2}(R^{\prime},\mathbf{\beta},Z)]\exp(-(R^{\prime})^2/2) \vphantom{T^{-1/2}} \right] + o(B_T/T) + O( | \tau^{2}_{1T} - \sigma^{\dag 2}_T |),
\end{split}
\end{align*}
\noindent uniformly on the Borel subsets on $\mathbb{R}$. As $R_{5,1}(R^{\prime},\mathbf{\beta},Z)$ is odd, $R_{5,1}(R^{\prime},\mathbf{\beta},Z)\exp(-(R^{\prime})^2/2)$ integrates to zero uniformly on the Borel subsets on $\mathbb{R}$. Therefore, we get an expansion of the form $\Upsilon^{\dag}_{Q,T} ( x ) = F _ { \mathcal{X} _ { r } ^ { 2 } } ( x ) + (B_T / T ) p _ { Q } ( x,\mathcal{K}_Q) f _ { \mathcal{X} _ { r } ^ { 2 } } ( x )$ such that $\sup_{y \in \mathbb{R}_+} | \int_{t \leq y} d\Upsilon^{\dag}_{Q,T} - \int_{t \leq y} dF_{\hat{Q}}| = o(B_T/T) + O( | \tau^{2}_{1T} - \sigma^{\dag 2}_T |)$. For an explicit method to determine the polynomials, see Remark 2.6 of \cite{chandra_valid_1979}. The arguments are the same for the bootstrap quadratic statistic $Q^{\ast}$, except that we have to replace the orders $o(B_T/T)$ by $o(1/T).$

\subsection*{Proof of Lemma \ref{Fourier_Edge_Bound}}

Let us define:
\begin{align}
\pi_T := T^{-1} \displaystyle\sum_{t=1}^{T} \displaystyle\sum_{j=1}^{T-B_T} \displaystyle\sum_{k=-B_T}^{B_T} k_{B_T}^{\ast}\left(k\right) \mathbb{E} \left[ g_t g_j g_ {j+k} \right],
\\
\mu_{3,T} := T^{2} \mathbb{E}\left[\bar{g}^3\right], \quad  \mu_{3,\infty} := \displaystyle\sum_{i=-\infty}^{\infty} \displaystyle\sum_{j=-\infty}^{\infty} \mathbb{E} \left[ g_0 g_{i} g_ {j}\right].
\end{align}
For computational convenience, let us consider the auxiliary signed measure $\Upsilon_{S,\infty}$, defined by its Fourier transform:
\begin{align*}
\mathcal{E}_\infty \left( \tau \right) &:= \int \exp \left( i\tau^\intercal x \right) d\Upsilon_{S,\infty}\left( x \right)
\\
&= \left\{ 1 + \frac{\mu_{3,\infty}}{\sqrt{T}\sigma^3_\infty} \left[ -\frac{1}{3} \left( i\tau \right)^3 - \frac{1}{2}  \left( i\tau \right)\right] \right\} \exp \left( -\frac{\tau^2}{2} \right).
\end{align*}

We can see that $\mathcal{E}_\infty$ differs from $\mathcal{E}^{\dag}_{T}$ and $\mathcal{E}^{\ast}_{T}$ by the variance and the third moment. Yet, we can control those differences by the following asymptotic results:

\begin{lemma}\label{original_aux}
Under Assumptions \ref{a1} --- \ref{Rosenblatt_mix}, we have: $\textnormal{(i) } \mu^\dag_{3,T} = \mu_{3,\infty} + O(T^{-1}), \textnormal{ (ii) } \pi^\dag_T = \mu_{3,\infty} + O(T^{-1}) \text{  and  \textnormal{(iii)} } \sigma^{\dag 2}_T = \sigma_\infty + o(1).$
\end{lemma}

Thus, by Esseen Lemma:
\begin{equation}
\sup_{x}\left\lvert\Upsilon_{S,\infty} \left( x \right) -  \Upsilon^{\dag}_{S,T} \left( x \right) \right\rvert \leq C \displaystyle\int_{\lvert\tau\rvert \leq T^{1/2+\epsilon}} \left\lvert\mathcal{E}_\infty \left( \tau \right) -  \mathcal{E}_T \left( \tau \right) \right\rvert \left\lvert \tau \right\rvert^{-1} d\tau + o\left(T^{-1/2}\right) + O(B_T^{-q}) + O(B_T/T).
\end{equation}
Furthermore, for the bootstrap statistic, we have the following asymptotic convergences of the variance and the third moment:
\begin{lemma}\label{FMB_aux}
Under Assumptions \ref{a1} --- \ref{Rosenblatt_mix}, we have: $\textnormal{(i) } \mu^{\ast}_{3,T} = \mu_{3,\infty} + O_p(T^{-1/2}) + O(B_T/T) + o(1) \text{  and \textnormal{(ii)} } \sigma^{\ast 2}_T = \sigma^2_\infty +  O_p(B_T^{-q}) + O_p((B_T/T)^{1/2}).$
\end{lemma}
Therefore, we have by the same argument: 
\begin{equation}
\sup_{x}\left\lvert\Upsilon_{S,\infty} \left( x \right) -  \Upsilon^{\ast}_{S,T} \left( x \right) \right\rvert \leq C \displaystyle\int_{\lvert\tau\rvert \leq T^{1/2+\epsilon}} \left\lvert\mathcal{E}_\infty \left( \tau \right) -  \hat{\mathcal{E}}_T \left( \tau \right) \right\rvert \left\lvert \tau \right\rvert^{-1} d\tau + o_p(T^{-1/2}).
\end{equation}

\subsection*{SM.4. Proof of Lemma \ref{original_aux}}

\textbf{(i)} First, $\pi_T = \mu_{3,T} + O(T^{-1})$ and $\pi^\dag_T = \mu^\dag_{3,T} + O(T^{-1})$ (\cite{gotze_second-order_1996}). Then, \begin{align*}
\lvert \pi_T - \pi^\dag_T \rvert &= \lvert T^{-1} \displaystyle\sum_{t=1}^{T} \displaystyle\sum_{j=1}^{T-l} \displaystyle\sum_{k=1}^{l} k_{B_T}^{\ast}\left(k\right) \mathbb{E} \left[ g_t\left(\beta_0\right) g_j\left(\beta_0\right) g_ {j+k}\left(\beta_0\right) \right]
\\
&- T^{-1} \displaystyle\sum_{t=1}^{T} w(t) \displaystyle\sum_{j=1}^{T-l} \displaystyle\sum_{k=1}^{l} k_{B_T}^{\ast}\left(k\right) \mathbb{E} \left[ g_t\left(\beta_0\right) g_j\left(\beta_0\right) g_ {j+k}\left(\beta_0\right) \right] \rvert
\\
&= T^{-1}\lvert \displaystyle\sum_{t=1}^{T}(1-w(t)) \displaystyle\sum_{j=1}^{T-l} \displaystyle\sum_{k=1}^{l} k_{B_T}^{\ast}\left(k\right) \mathbb{E} \left[ g_t\left(\beta_0\right) g_j\left(\beta_0\right) g_ {j+k}\left(\beta_0\right) \right] \rvert
\\
&= O(B_T/T).
\end{align*}
Thus, $\lvert \mu_{3,T} - \mu^\dag_{3,T} \rvert \leq \lvert \mu_{3,T} - \pi_T \rvert + \lvert \pi_T - \pi^{\dag}_T \rvert + \lvert \pi^{\dag}_T - \mu^\dag_{3,T} \rvert = O(T^{-1}) + O(B_T/T) + O(T^{-1})$. As $\mu_{3,T}$ tends to $\mu_{3,\infty}$ by definition, the proof is concluded.

\textbf{(ii)} Recall that $\lvert \mu_{3,T} - \mu^\dag_{3,T} \rvert = O(B_T/T)$ and $\pi^{\dag}_T = \mu^\dag_{3,T} + O(T^{-1})$. Then, the lemma follows by triangular inequality, and as $\mu_{3,T}$ tends to $\mu_{3,\infty}$ by definition.

\textbf{(iii)} $\sigma^{\dag 2}_T = \Var\left[T^{1/2}\bar{g}_T\right] = \sigma_\infty + o(1)\text{ by definition.}$

\subsection*{SM.5. Proof of Lemma \ref{FMB_aux}}

\textbf{(i)} We have:
\begin{align*}
&\mu^{\ast}_{3,T} = T^{1/2} \mathbb{E}^{\ast} \left[ T^{1/2}\bar{g}_{T}^{\ast 3} \right]=T^{-1} \displaystyle \sum_{r=1}^{T} \displaystyle \sum_{s=1}^{T} \displaystyle \sum_{t=1}^{T} \mathbb{E}^{\ast}\left[ g_{T,r}^{\ast} g_{T,s}^{\ast} g_{T,t}^{\ast} \right] \\
&= T^{-1} \displaystyle \sum_{t=1}^{T} \mathbb{E}^{\ast}\left[ g_{T,t}^{\ast 3} \right] = T^{-1}\displaystyle \sum_{t=1}^{T}g_{T,t}^3 = \mathbb{E}\left[ T^{-1}\displaystyle \sum_{t=1}^{T} g_{T,t}^{3} \right] + O_p(T^{-1/2}).
\end{align*}
Now, $\mathbb{E}\left[ T^{-1}\displaystyle \sum_{t=1}^{T}g_{T,t}^{3} \right] = \mathbb{E}\left[g_{T,s_1}^{3}\right] + O(B_T/T)$, where $s_1 := \left[\frac{T}{2}\right]$. Thus, $\mu^{\ast}_{3,T} = \mathbb{E}\left[g_{T,s_1}^{3}\right] + O_p(T^{-1/2}) + O(B_T/T)$. Then, considering that $g_{T,s_1}$ is a standardized weighted sum and from any CLT for strongly mixing process under Lindeberg conditions (see e.g.\ \cite{rio_about_1997}), $\mathbb{E}\left[g_{T,s_1}^{3}\right] = \mu_{3,\infty} + o(1)$.

\textbf{(ii)} Recall that $\sigma^{\ast 2}_T = \mathbb{E}^{\ast} \left[ T\bar{g}_{T}^{\ast 2} \right] = T^{-1}\displaystyle \sum_{t=1}^{T}g_{T,t}^2$. Thus, $\sigma^{\ast 2}_T$ has the form of an automatically positive semi-definite HAC estimator (\cite{smith_automatic_2005}). From the properties of HAC-type estimators (see Section \ref{sectionHAC}), we have: $\sigma^{\ast 2}_T = \sigma^2_\infty + O_p(B_T^{-q}) + O_p((B_T/T)^{1/2})$, where $q$ is the Parzen exponent. As a consequence, the bootstrap variance converges to the long-run one.

\subsection*{SM.6. Proof of Corollary \ref{HOtildeQ}}

From the Taylor expansion defining $\tilde{Q}$ in \eqref{tildeQStat} we have $\tilde{Q}(\beta_0)=\hat{Q}(\beta_0)+R_T,$ where $\mathbb{P}[R_T > \delta_T] = \delta_T$ for some positive sequence $\delta_T = o(T^{-1}).$ It follows that:
\begin{align*}
\mathbb{P}[\tilde{Q}(\beta_0) \leq x] &= \mathbb{P}[\hat{Q}(\beta_0) + R_T \leq x] \\
&= \mathbb{P}[\hat{Q}(\beta_0) + R_T \leq x \big| |R_T| \leq \delta_T] \mathbb{P}[|R_T| \leq \delta_T]  \\
&+ \mathbb{P}[\hat{Q}(\beta_0) + R_T \leq x \big| |R_T| > \delta_T]\mathbb{P}[|R_T| > \delta_T] \\
&= \mathbb{P}[\hat{Q}(\beta_0) + R_T \leq x \big| |R_T| \leq \delta_T] (1-\delta_T) \\
&+ \mathbb{P}[\hat{Q}(\beta_0) + R_T \leq x \big| |R_T| > \delta_T]\delta_T \\
&= \mathbb{P}[\hat{Q}(\beta_0) + R_T \leq x \big| |R_T| \leq \delta_T] + C\delta_T \\
&\geq \mathbb{P}[\hat{Q}(\beta_0) \leq x-\delta_T] + C\delta_T.
\end{align*}
\noindent Following the same argument, we have:
\begin{align*}
\mathbb{P}[\tilde{Q}(\beta_0) \leq x] &= \mathbb{P}[\hat{Q}(\beta_0) + R_T \leq x \big| |R_T| \leq \delta_T] + C\delta_T \\
&\leq \mathbb{P}[\hat{Q}(\beta_0) \leq x+\delta_T] + C\delta_T.
\end{align*}
As both $\mathbb{P}[\hat{Q}(\beta_0) \leq x]$ and $\mathbb{P}[\tilde{Q}(\beta_0) \leq x]$ are bounded above and below up to the $C\delta_T = o(T^{-1})$ term, we get: 
\begin{align}
\sup_{x \in \mathbb{R}^{+}}|\mathbb{P}[\tilde{Q}(\beta_0) \leq x] - \mathbb{P}[\hat{Q}(\beta_0) \leq x]| &\leq \sup_{x \in \mathbb{R}^{+}}|\mathbb{P}[\hat{Q}(\beta_0) \leq x-\delta_T] - \mathbb{P}[\hat{Q}(\beta_0) \leq x + \delta_T] + o(T^{-1})| \\
&\leq \sup_{x \in \mathbb{R}^{+}}| \mathbb{P}[\hat{Q}(\beta_0) \leq x-\delta_T] - \mathbb{P}[\hat{Q}(\beta_0) \leq x + \delta_T]| + o(T^{-1}) \\
&= o(T^{-1}),
\end{align}
\noindent by continuity of the probability distribution. Then, the result follows from an application of Theorem \ref{higher_order}.

\subsection*{SM.7. CPU time}

\begin{table}[htbp]
\begin{center}
\caption{Comparison of CPU time (in seconds) between FMB and MBB.}
\begin{tabular}{l l l l}
\hline
\hline

	\\
	\multicolumn{4}{c}{$ACD(1,0)$} \\	
	&&&\\
  &&$B_T=3$ & $B_T=5$ \\
		&&&\\
  \multirow{1}{*}{$T=250$}
	& $\hat{Q}_{FMB}$ & 0.72   & 0.65 \\ 
  & $W_{MBB}$  & 362.61 & 286.29 \\ 
	\\
	
   \multirow{1}{*}{$T=500$}
	& $\hat{Q}_{FMB}$ & 0.75   & 0.68 \\ 
  & $W_{MBB}$  & 617.46 & 467.86 \\ 
	&&&\\ 
	\\
	\multicolumn{4}{c}{$ACD(1,1)$} \\
	&&&\\
	 &&$B_T=3$ & $B_T=5$ \\
		&&&\\
  \multirow{1}{*}{$T=250$}
	
	& $\hat{Q}_{FMB}$ & 0.74  & 0.74 \\ 
  & $W_{MBB}$  & 1518.59 & 1104.58 \\ 
	\\
	
   \multirow{1}{*}{$T=500$}
	
	& $\hat{Q}_{FMB}$ & 0.78  & 1.21 \\ 
  & $W_{MBB}$  & 2963.06 & 1751.53 \\ 
	\\
  \hline
	\hline
\end{tabular}
\label{Tab: CPU}
\caption*{The number of bootstrap samples is $R=2500$ for both methods.}
\end{center}
\end{table}

\newpage
\subsection*{SM.8. Coverages for CR of ACD(1,0) and ACD(1,1) with $T=150.$}

The results we show in Table \ref{Tab: T150} must be interpreted with caution since more than $10\%$ of the estimators did not converge for this sample size in the Monte Carlo simulations.

\begin{table}[htbp]
\begin{center}
\caption{Coverage of CR, a comparison between first and higher-order correct methods.}
\begin{tabular}{l l l l l l l l}
\hline
\hline

	\\
	
	\multicolumn{8}{c}{ACD(1,0)}\\
	
  && \multicolumn{3}{c}{ $B_T=3$ } & \multicolumn{3}{c}{$B_T=5$} \\
	
  \multicolumn{2}{l}{Coverages:}  & 0.90 & 0.95 & 0.99   & 0.90 & 0.95 & 0.99 \\
	\\
  \multirow{6}{*}{\rotatebox[origin=c]{90}{$T=150$}} & $\hat{Q}_{FMB}$ & 0.88 & 0.92 & 0.96  & 0.85 & 0.89 & 0.94 \\ 

  &$\tilde{Q}_{FMB}$ & 0.89 & 0.92 & 0.96 & 0.88 & 0.91 & 0.96 \\
	&$\tilde{Q}_{FMB,2}$  & 0.88 & 0.92 & 0.97 & 0.85 & 0.90 & 0.96  \\
	&$\hat{Q}_{\mathcal{X}^2_r}$ & 0.85 & 0.9 & 0.94 & 0.84 & 0.87 & 0.93 \\ 
  &$W_{MBB}$  & 0.9 & 0.94 & 1.00 & 0.92 & 0.96 & 1.00 \\ 
  &$W_{\mathcal{X}^2_{p}}$  & 0.78 & 0.85 & 0.92 & 0.74 & 0.80 & 0.88  \\ 
	
	&&&&&&&\\
	
	\multicolumn{8}{c}{ACD(1,1)}\\
	
	 && \multicolumn{3}{c}{ $B_T=3$ } & \multicolumn{3}{c}{$B_T=5$} \\
	 \multirow{6}{*}{\rotatebox[origin=c]{90}{$T=150$}} & $\hat{Q}_{FMB}$ & 0.87 & 0.92 & 0.96 & 0.83 & 0.88 & 0.93\\ 

  &$\tilde{Q}_{FMB}$ & 0.87 & 0.9 & 0.94 & 0.84 & 0.88 & 0.91\\
	&$\tilde{Q}_{FMB,2}$   & 0.78 & 0.84 & 0.91 & 0.72 & 0.79 & 0.88  \\
	&$\hat{Q}_{\mathcal{X}^2_r}$ & 0.83 & 0.88 & 0.93 & 0.80 & 0.85 & 0.90\\ 
  &$W_{MBB}$ & 0.95 & 0.98 & 1.00 & 0.97 & 0.99 & 1.00 \\ 
  &$W_{\mathcal{X}^2_{p}}$  & 0.67 & 0.73 & 0.83 & 0.62 & 0.69 & 0.79\\ 
	\\
  \hline
	\hline
\end{tabular}
\label{Tab: T150}
\caption*{

The true values of the unknown parameters of the ACD(1,0) and ACD(1,1) are $\omega = 1.5$ and $\beta_1 = 0.25$ and $\beta_2 = 0.25$.}
\end{center}
\end{table}

\subsection*{SM.9. Graphical illustration of the use of $\tilde{Q}$ as in Remark \ref{nomono}.}

\begin{figure}[H]
\begin{center}
\caption{Graphical illustration of Remark \ref{nomono}.}\label{nomonoC}
\includegraphics[width = 15cm, height = 10cm]{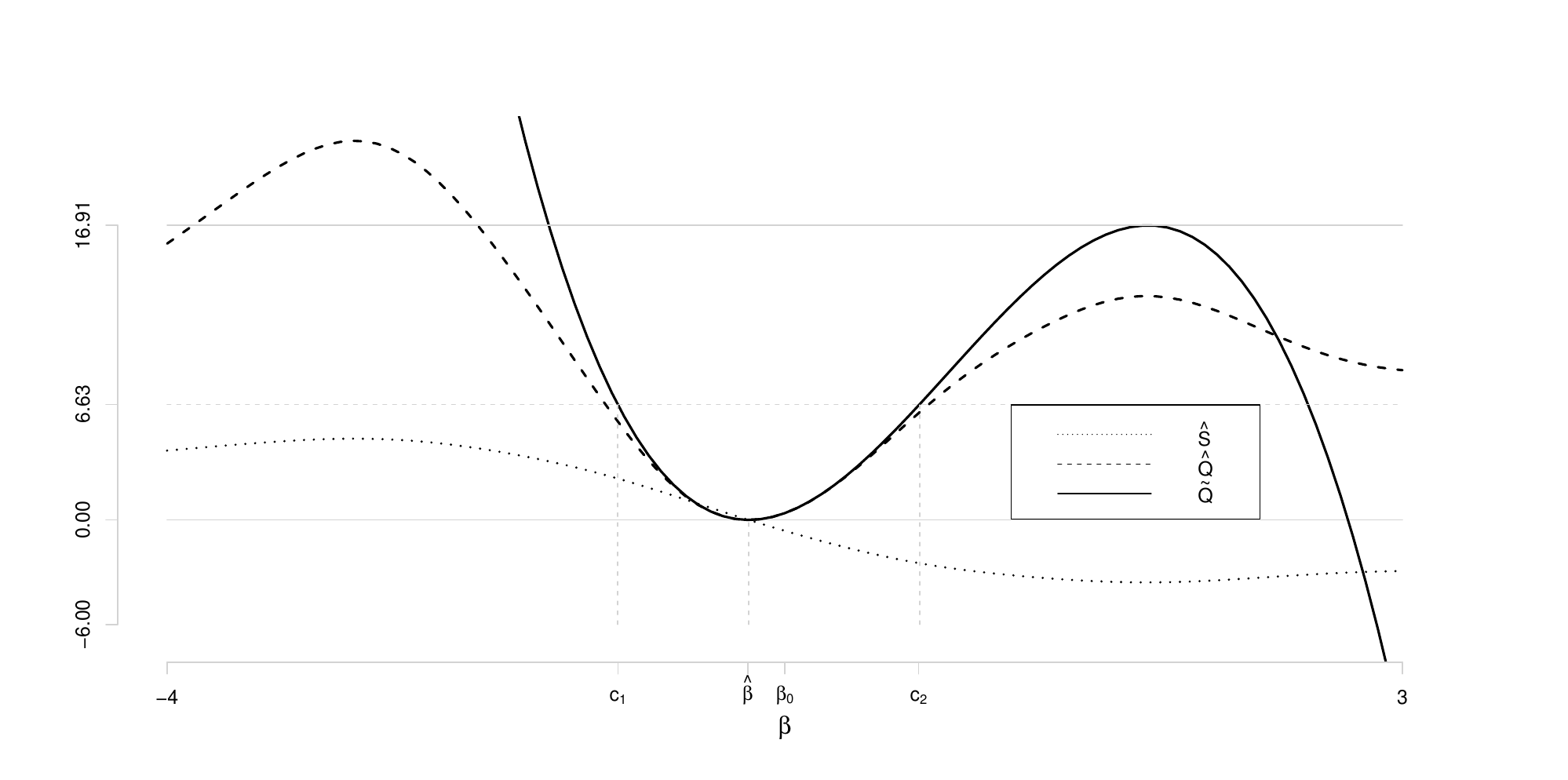}
\caption*{The studentized score $\hat{S}(\beta_0)$ of the Cauchy location parameter $\beta_0$, its square $\hat{Q}(\beta_0)$ and the approximation by $\tilde{Q}(\beta_0)$. In this example, the bounds of the CI are be $c_1=-0.81$ and $c_2=0.65.$}
\end{center}
\end{figure}

To illustrate the behavior of the approximation $\tilde{Q}$, we consider the problem of estimating the location parameter $\beta_0$ of a Cauchy distribution, from an i.i.d.\ sample of size $T=30.$ This example does not correspond to our inferential setting of interest because of its lack of finite moments, but it has the advantage to provide a clear picture of the worst-case scenario, where the studentized score function $\hat{S}(\beta_0)$ fails to be one-to-one in $\beta_0$, making the direct inversion technique impossible. Figure \ref{nomonoC} shows the plot of $\hat{S},$ $\hat{Q},$ and its approximation $\tilde{Q}.$ The local maximum of $\tilde{Q}$ ($16.91$ for this sample) is the maximum level allowing to define a simply connected CI as level set. As discussed in Remark \ref{nomono}, this maximum increases proportionally to the sample size $T,$ by definition of $\tilde{Q}$ as a cubic polynomial. In this example, we draw a line at $6.63$ --- the 99\% percentile of a chisquare distribution with one degree of freedom --- to illustrate the applicability of $\tilde{Q}$ already with $T=30.$

\subsection*{SM.10. Verification of Assumptions \ref{a1}---\ref{EE5} for the AR($p$) example.}

As an example of process where Assumptions \ref{a1}---\ref{EE5} are satisfied, we take the OLS moment indicators for the autoregressive parameters of an AR($p$) process $Y_t = \sum_{k=1}^{p} \theta_k Y_{t-k} + e_t = \sum_{j=0}^\infty w_j e_{t-j}$, with $e_t \overset{i.i.d.} \sim \mathcal{N}(0,\sigma^2)$ and $\lvert w_j \rvert \leq \delta^{-1} \exp(- \delta j).$ Namely, $g_t = g(Y_t,...,Y_{t-p};\theta)$ is the $p-$dimensional moment condition vector with entries $g_{i}(Y_t,...,Y_{t-p};\theta) = Y_{t-i}(Y_t-\sum_{k=1}^p \theta_k Y_{t-k}),$ $\forall i=1,...,p.$ We define the sub-sigma-fields $D_t:= \sigma\langle e_t,...,e_{t-p} \rangle.$ The innovations $e_t$ being i.i.d.\ Gaussian, the verification of Assumptions \ref{a1} and \ref{a2} follows from the orthogonality conditions of OLS and from the existence of all the moments. We verify the exponential decay in Assumptions \ref{Rosenblatt_mix} and \ref{Markov_type} as the strongly mixing coefficient of Assumption \ref{Rosenblatt_mix} and the Markov-type condition of Assumption \ref{Markov_type} drop to zero for independent random variables. For Assumption \ref{RV_approx}, there exists a constant $\delta > 0$ such that for $t,m = 1,2,...$ and $m > \delta^{-1}$, we can approximate $Y_{t}$ by a $\mathcal{D}_{t-m}^{t+m}$-measurable random vector $Y_{t,m}^{\ddagger}$, such that $\mathbb{E}\left\| Y_t - Y_{t,m}^{\ddagger} \right\| \leq \delta^{-1} \exp\left(-\delta m\right),$ for instance by taking $Y_{t,m}^{\ddagger} = \sum_{j=0}^m w_j e_{t-j}.$ Then, $\mathbb{E}\left\| g(Y_t,...,Y_{t-p};\theta) - g(Y_{t,m}^{\ddagger},...,Y_{t-p,m}^{\ddagger};\theta) \right\| \leq \sum_{i=1}^p \mathbb{E} \lvert g_{i}(Y_t,...,Y_{t-p};\theta) - g_{i}(Y_{t,m}^{\ddagger},...,Y_{t-p,m}^{\ddagger};\theta)\rvert.$ Therefore, it is sufficient to show that each element of the latter sum is decreasing exponentially with $m,$ to get a process $g^{\ddagger}_{t,m} = g(Y_{t,m}^{\ddagger},...,Y_{t-p,m}^{\ddagger};\theta)$ approximating $g_t$ with exponentially decaying error, as required by Assumption \ref{RV_approx}. Using a Taylor expansion, we can write $g_{i}(Y_t,...,Y_{t-p};\theta) - g_{i}(Y_{t,m}^{\ddagger},...,Y_{t-p,m}^{\ddagger};\theta) = Z_i^{\intercal} D_t^{\ddagger} + D_t^{\ddagger \intercal}A D_t^{\ddagger},$ $\forall i=1,...,p,$ where $Z_i$ is a vector of Gaussian random variables, $A$ is a matrix of deterministic constants and $D_t^{\ddagger}$ is a vector with entries $Y_{t-i}-Y_{t-i,m}^{\ddagger},$ $\forall i=1,...,p.$ Then, using Cauchy-Schwarz inequality on $Z_i^{\intercal} D_t^{\ddagger}$ allows to verify directly Assumption \ref{RV_approx} on ${g_t}$ by making use of the exponential decay of each element in $D_t^{\ddagger}$. We can verify Assumptions \ref{Condition_Cramer} and \ref{EE5} by using the Riemann-Lebesgue lemma and the continuity of the function $g,$ $g_t$ having always a density.

\end{document}